\documentclass[12pt]{spieman}  
\usepackage{amsmath,amsfonts,amssymb}
\usepackage{graphicx}
\usepackage{setspace}
\usepackage{tocloft}
\usepackage{array}

\usepackage[margin=1in]{geometry}
\usepackage{graphicx}
\usepackage{amsmath}
\usepackage{amssymb}
\usepackage{booktabs}
\usepackage{hyperref}
\usepackage[numbers]{natbib}
\usepackage{xcolor}
\usepackage{caption}
\usepackage{tikz}
\usetikzlibrary{arrows.meta,decorations.pathreplacing}
\usepackage{url}

\title{An RFSoC-Based 100 GbE Spectrometer/Correlator for the Baryon Mapping Experiment: Digital Backend Design and On-Sky Validation}

\author[a*]{Liju Philip}
\author[a]{Paul O'Connor}
\author[a]{Benjamin Saliwanchik}
\author[a]{An{\v z}e Slosar}
\author[a]{Emi Tamura}
\affil[a]{Brookhaven National Laboratory, Instrumentation Division, Bldg 535, Technology St., Upton, NY 11973, USA}

\newcommand{\cygA}{Cygnus~A}
\newcommand{\kms}{\ensuremath{\mathrm{km\,s^{-1}}}}

\newcommand{\HI}{H\textsc{i}}

\cftpagenumbersoff{figure}
\cftpagenumbersoff{table}
\begin{document}
\maketitle

\begin{abstract}
We describe the implementation and on-sky commissioning of a two-input spectrometer/correlator built on a Xilinx ZCU208 Radio Frequency System-on-Chip (RFSoC) with a 100\,Gigabit Ethernet (GbE) visibility output, deployed on the Baryon Mapping eXperiment (BMX) interferometric array at Brookhaven National Laboratory (BNL). A next-generation BMX backend depends on two capabilities that have not previously been demonstrated together on this instrument: integrating the RFSoC's direct-sampling data converters and its programmable logic into a working correlator, and sustaining lossless visibility transport over a 100\,GbE link.
Two inputs are digitized directly at 1050\,MHz, with no analog downconversion, and each is channelized into 2048 frequency channels. A correlator forms the two autocorrelations and the real and imaginary parts of the cross-correlation, accumulating each in on-chip memory, and a transmitter reads the accumulated spectra out as User Datagram Protocol (UDP) packets at a fixed, firmware-controlled rate that keeps the 100\,GbE core from being oversubscribed. Two independent campaigns, at 500\,ms and 100\,ms accumulation, returned 958{,}918 consecutive accumulations with no packet lost, duplicated, or malformed, and a Cauchy--Schwarz bound holds across all $1.96\times10^{9}$ channel-samples, confirming that the transported fields are correctly identified and scaled.

Three sky signals confirm that the visibilities are physically sensible: transits of compact sources in six Global Navigation Satellite System (GNSS) bands, showing the expected fringe structure; the \cygA{} zenith transit, detected coherently in cross-correlation at a time consistent with prediction in both campaigns; and a Galactic \HI{} plane crossing, detected in autocorrelation and reproduced independently on each dish. The correlator arithmetic is parameterizable in the number of inputs, growing quadratically with it, and the RF front end is already provisioned for BMX's full dual-polarization array of eight inputs. Scaling to eight inputs is analyzed against an occupancy model validated in simulation of the deployed firmware: with the visibility buffer moved to UltraRAM (URAM), the buffering fits the device, while multi-tile synchronization, logic utilization, and timing closure remain to be established by an implementation run, so this two-input system is a commissioned foundation for that extension rather than a demonstration of it.
\end{abstract}


{\noindent \footnotesize\textbf{*}Liju Philip,  \linkable{lphilip@bnl.gov} }

\begin{spacing}{2}   

\section{Introduction}
\label{sect:intro}
A conventional radio-astronomy backend is a chain of separate stages: analog downconversion to an intermediate frequency, digitization on dedicated converter boards, and transport of raw samples to wherever
the channelization and correlation are performed. 
A Radio Frequency System-on-Chip (RFSoC) removes most of that chain by placing the data converters on the same die as programmable logic. The work reported here uses a Xilinx ZCU208 evaluation platform\footnote{\url{https://docs.amd.com/v/u/en-US/ug1410-zcu208-eval-bd}}, which carries an XCZU48DR Zynq UltraScale+ RFSoC with a CLK104\footnote{\url{https://docs.amd.com/r/en-US/ug1437-clk104/}} RF clocking card and a zSFP+ optical cage for the 100 Gigabit Ethernet (GbE) link. The XCZU48DR integrates eight 14-bit analog-to-digital converters (ADCs) alongside a field-programmable gate array (FPGA) fabric and two hardened 100\,GbE media-access controllers (MACs), so the observing band is sampled directly and the samples are channelized, correlated and accumulated in the adjacent fabric, leaving the device only as reduced visibilities.
Direct RF sampling with these parts has been characterized for radio-astronomy use by Liu et al.\cite{liu2021,liu2023}, and the Collaboration for Astronomy Signal Processing and Electronics Research (CASPER\footnote{\url{https://casper-astro.github.io/}}) has developed an open tool flow for building instruments on them \cite{hickish2016}.

For an interferometer measuring the 21\,cm \HI{} line, direct sampling on an RFSoC shortens the analog chain ahead of each antenna, reducing the stages in which spurious products and gain drift can arise, and gives every converter a common sample clock with a fixed phase relationship, without which the relative phase between array elements would drift and the fringe would decohere. Both are binding constraints for the compact transit arrays that 21\,cm intensity mapping relies on.

The Baryon Mapping eXperiment (BMX) telescope is one such array: a zenith-pointing drift-scan radio interferometer at Brookhaven National Laboratory (BNL, $40.87^\circ$\,N, $72.87^\circ$\,W), originally described by O'Connor et al.\cite{oconnor2020}. It is a testbed for post-reionization 21\,cm intensity mapping -- tracing large-scale structure through redshifted \HI{} emission rather than resolving individual galaxies.
The array comprises four 4\,m offset parabolic dishes (North/South/East/West) on $8.8$\,m North--South and East--West baselines, with a 1.1--1.55\,GHz band covering $0<z<0.3$ and a system temperature of $T_{\rm sys}\approx70$\,K. Such a measurement depends less on raw sensitivity than on spectral cleanliness and long-term stability, both of which the shorter analog chain and the shared sample clock address directly. The match extends to scale: BMX's four dual-polarization dishes produce eight signals, exactly the converter count of one XCZU48DR, so an array that presently requires multiple digitizer boards could in principle be
read out by a single device. O'Connor et al.\cite{oconnor2020} reported $\sim$7000\,hours of data (January 2019--May 2020), validated with Global Navigation Satellite System (GNSS) satellites, Galactic \HI{}, and repeated daily \cygA{} transits, and explicitly identified a compact RFSoC-based digitizer/F-engine as the planned next-generation backend.

This paper reports a first step toward that upgrade: a two-input implementation built to establish feasibility rather than to produce science. The questions are whether the RFSoC's data converters and digital signal processing (DSP) fabric can be integrated into a correlator that runs on sky, and whether visibility data can be carried off the device over 100\,GbE without loss. We report both the digital design (Section~\ref{sec:backend}) and its validation on real sky data (Sections~\ref{sec:observations}--\ref{sec:onsky}). Two independent engineering test campaigns are used, three days apart and at different accumulation times (500\,ms and 100\,ms); we treat the array's physical baseline, beam size, and system temperature as known quantities from O'Connor et al.\cite{oconnor2020} rather than re-deriving them. The sky data are used as validation signals, not as measurements: no absolute flux or brightness-temperature calibration is applied anywhere in this paper, and no cosmological or astrophysical result is claimed. The focus is accordingly on (i) the backend implementation itself, (ii) confirming, with three independent on-sky signals, that it produces complete, physically sensible visibilities, and (iii) assessing what scaling this design to BMX's full dual-polarization array of eight inputs would require.

The paper is organized as follows. Section~\ref{sec:sys_overview} gives an overview of the system, outlining
the signal path from antenna to the backend computer. Section~\ref{sec:backend} details the digital backend: platform and tool flow, RF data conversion, channelization, correlation and accumulation, timing and synchronization, visibility packetization, the packetizer's frame state machine and flow-control design (including a wire-format corruption issue and its resolution), and its pre-deployment verification. Section~\ref{sec:capture} describes data capture and streaming validation. Section~\ref{sec:observations} summarizes the observing campaigns used. Section~\ref{sec:onsky} presents on-sky validation via GNSS transits, the \cygA{} transit, and a Galactic \HI{} plane crossing. Section~\ref{sec:discussion} discusses the two-input system's known limitations and what scaling it to eight inputs would require. We conclude in Section~\ref{sec:conclusion}.

\section{System overview}
\label{sec:sys_overview}
\begin{figure}[htbp]
\centering
\includegraphics[width=0.8\linewidth]{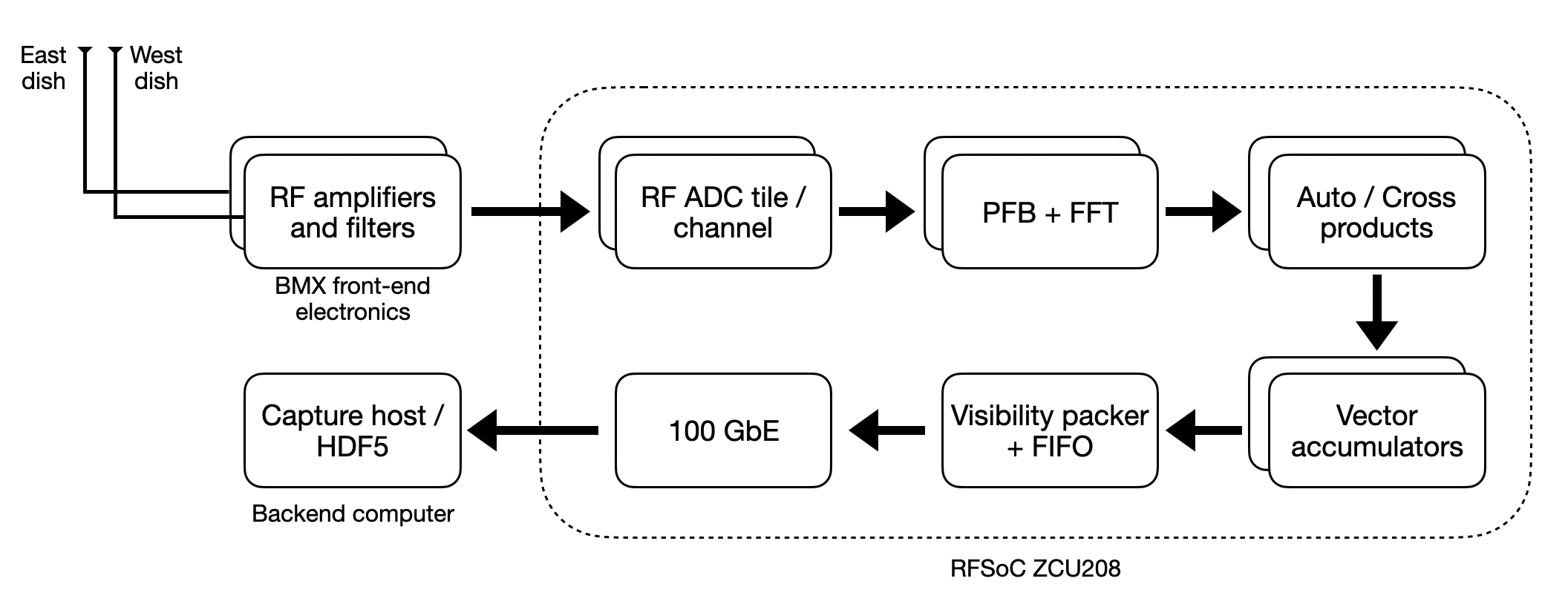}
\caption{Signal-path block diagram, dish to archived HDF5. The dashed boundary encloses the RFSoC. Each stage is described in Sections~\ref{sec:backend}--\ref{sec:capture}.}
\label{fig:blockdiagram}
\end{figure}

Figure~\ref{fig:blockdiagram} summarizes the signal path described in Sections~\ref{sec:backend}--\ref{sec:capture}: each dish and feed is followed by analog filtering and gain, an RF-ADC tile channel, a polyphase filter bank (PFB) and fast Fourier transform (FFT) channelizer, the auto/cross correlator, per-channel vector accumulators, the visibility packer and its first-in-first-out memory (FIFO), the 100\,GbE link, and the capture host writing Hierarchical Data Format 5 (HDF5) files. The dashed boundary in the figure
marks what is on the RFSoC: everything from digitization to packet transmission runs in a single device, and the only components outside it are the array's existing analog front end and the capture computer.

The analog front end is inherited unchanged from the existing instrument described by O'Connor et al.\cite{oconnor2020}. Each feed's orthomode transducer drives a temperature-stabilized amplifier and filter chain that defines the 1.1--1.55\,GHz passband. No mixing or downconversion occurs anywhere in the chain: the band reaches the converters at its observed frequency and is sampled directly in the third Nyquist zone (Section~\ref{sec:rfdc}).

Sampling takes place in the RF Data Converter (RFDC) block, the hardened converter subsystem on the RFSoC die that contains the RF-ADC tiles and presents their output to the programmable logic as an AXI4-Stream (Advanced eXtensible Interface 4 Stream) interface. Each digitized stream is then channelized by a PFB feeding an FFT, the standard arrangement for radio spectrometers, which divides the band into frequency channels with
better isolation between them than an FFT alone provides. The channelized streams pass to the correlator, which forms the autocorrelation of each input and the cross-correlation between them, and these products are integrated in vector accumulators -- one running sum per frequency channel -- held in on-chip block random access memory (BRAM). At the end of each integration the accumulated spectra are gathered by the visibility packer into fixed-width words, buffered in a FIFO, and transmitted as User Datagram Protocol (UDP) frames over the 100\,GbE link to the capture host, which assembles them into complete spectra and writes them to HDF5 files. Section~\ref{sec:backend} describes each of these stages in detail, and Section~\ref{sec:capture} the capture and archiving chain.

Table~\ref{tab:backendcomp} compares the new backend with the graphics processing unit (GPU)-based backend it replaces \cite{oconnor2020}. The main gains are in digitizer resolution (14 rather than 8 bits), a sample clock synthesized on the board within the converters' rating rather than by underclocking commercial digitizer cards, a register-programmable integration time (100\,ms and 500\,ms used here, against the $\sim$1\,s used previously), and the consolidation of digitization, channelization, correlation and packetization onto a single device whose output stream was verified to be loss-free over both campaigns (Section~\ref{sec:capture}). The previous backend retains capabilities not yet reproduced here: eight correlated inputs with twelve
like-polarization cross-correlations, a fine-resolution zoom band around the 21\,cm line, and online radio-frequency interference (RFI) clipping. A 21\,cm zoom spectrometer will be implemented on the RFSoC in future firmware; it is beyond the
scope of this paper, whose aim is to demonstrate that an RFSoC backend is feasible for BMX.

\begin{table}[htbp]
\centering
\caption{Comparison between the previous BMX digital backend and the RFSoC/100\,GbE backend of this work. Values for the previous backend are as stated in O'Connor et al.\cite{oconnor2020}.}
\label{tab:backendcomp}
\small
\begin{tabular}{>{\raggedright\arraybackslash}p{0.22\linewidth}>{\raggedright\arraybackslash}p{0.35\linewidth}>{\raggedright\arraybackslash}p{0.35\linewidth}}
\toprule
 & Previous backend (2020) & RFSoC/100\,GbE (this work) \\
\midrule
Digitizer & Spectrum M4-2223-x8 PCIe cards, 2 inputs per card, 4 cards in 2 PCs
  & ZCU208 RFSoC with 8 RF-ADCs \\
ADC resolution & 8 bit & 14 bit \\
Sample clock & 1.1\,GS/s external clock; cards rated for 1.25\,GS/s, underclocked
  & 1.05\,GS/s from on-board LMK04828 reference and tile PLLs \\
Band (3rd Nyq.\ zone) & 1.1--1.55\,GHz & 1.05--1.575\,GHz \\
Data path to F-engine & PCIe x8 Gen\,2 to host memory, then to GPU; bus bandwidth limits inputs per card
  & on-chip AXI4-Stream, no host bus \\
F-engine & GPU (GTX 1050\,Ti), CUDA; long FFT binned to channels
  & FPGA fabric; 4-tap PFB + 4096-point real FFT \\
Channels, full band & 2048 & 2048 \\
21\,cm zoom band & yes (finer grid around the line) & planned; not in this work \\
Arithmetic & 8-bit samples converted to 32-bit float & fixed point; 18-bit FFT, 64-bit integer accumulators \\
Correlated inputs & 8 (4 dishes $\times$ 2 pol) & 2 (single pol) \\
Products & 8 auto + 12 like-pol cross & 2 auto + 1 cross \\
Integration time & $\approx$1\,s used (32 $\times$ 30\,ms chunks; chunk length configurable) & register-programmable; 500\,ms and 100\,ms used \\
Online RFI excision & $>4\sigma$ median clip ($\sim$1\% removed) & none; raw visibilities recorded \\
Output transport & host disk & 100\,GbE UDP to separate capture host \\
Data volume & $\sim$24\,GB/day & 11.5 / 57.5\,GB/day (500 / 100\,ms) \\
\bottomrule
\end{tabular}
\end{table}

\section{The RFSoC/\texorpdfstring{100\,GbE}{100 GbE} Digital Backend}
\label{sec:backend}

\subsection{Platform and tool flow}
\label{sec:platform}
The backend is implemented on the ZCU208 platform introduced in Section~\ref{sect:intro}. For this work, we use two RF-ADC channels out of the available eight (Section~\ref{sec:rfdc}), and one of the two 100\,GbE MACs drives the output link through the zSFP+ cage, whose four lanes are combined by a breakout cable into the single 100\,GbE connection to the capture host.

The firmware used in this work is built in the CASPER MATLAB/Simulink/System Generator tool flow, in which the design is drawn as a block diagram of parameterized DSP primitives -- RFDC, PFBs, FFTs, accumulators, and 100\,GbE interfaces -- rather than written in a hardware description language (HDL). Block parameters such as transform length, tap count, and arithmetic precision are set at design capture and propagate through the generated logic, so the channel count, accumulator format, and correlator width reported here are properties
of the model rather than of hand-written code. This is what makes the scaling analysis of Section~\ref{sec:scaling} a question of re-parameterization and resource budget rather than of redesign. The model is compiled through Vivado (2025.1 for the deployed build) to a bitstream. Run-time control uses the \texttt{casperfpga}\footnote{\url{https://github.com/casper-astro/casperfpga}} Python API.

Two networks are used, and their separation is architecturally important. A 1\,GbE control-plane link carries register reads and writes -- accumulation length, reset strobes, and status counters -- while the 100\,GbE data plane carries only streamed visibility payload, as one-way UDP with no return path. All instrument state and telemetry are therefore out-of-band with respect to the science data, and no control traffic can perturb the timing of the visibility stream.

\subsection{RF data converter}
\label{sec:rfdc}
The RF-ADC tiles for this work are configured for direct sampling with no on-chip decimation and no digital mixing: decimation factor $1\times$, mixer type bypassed, real-to-real mixer mode, and real digital output. The sample rate is
\begin{equation}
  f_s = 1050\,\mathrm{MHz},
\end{equation}
placing the third Nyquist zone at 1050--1575\,MHz and covering the BMX analog passband of 1100--1550\,MHz with margin at both edges. The analog bandpass filter ahead of each ADC, whose stopband attenuation is 70\,dB \cite{oconnor2020}, rejects signal power from all other Nyquist zones before digitization, so no separate digital alias-suppression step is required downstream.

Background calibration is enabled, and the RF-ADC calibration mode is set to Mode 2. The mode selection depends on where the signal falls within the first Nyquist band after aliasing: Mode 1 is specified for $0.4f_s$ to $f_s/2$ and Mode 2 for 0 to $0.4f_s$\footnote{\url{https://docs.amd.com/r/en-US/pg269-rf-data-converter/Calibration-Modes}}.
Sampling in the third Nyquist zone folds the band down by $f_s$, so the 1100--1550\,MHz passband aliases to 50--500\,MHz, of which 50--420\,MHz lies within the Mode 2 range; the upper $\sim$80\,MHz, corresponding to RF 1470--1550\,MHz, falls above the $0.4f_s = 420$\,MHz boundary and is nominally Mode 1 territory. Mode 2 was selected as the better match to the bulk of the occupied band.

Each converter delivers four real samples per fabric clock, so the AXI4-Stream fabric clock is
\begin{equation}
  f_{\rm clk} = \frac{f_s}{4} = 262.5\,\mathrm{MHz},
\end{equation}
a clock period of 3.810\,ns, and all downstream logic is synchronous to it.

Tile Phase-Locked Loops (PLLs) are enabled and locked to a 250\,MHz reference generated by the LMK04828 jitter cleaner on the ZCU208's CLK104 RF clocking card and routed to the RF-ADC tile reference inputs. The same device supplies the 125\,MHz programmable-logic clock and a 5\,MHz SYSREF (required for multi-tile synchronization, which is not engaged in this work). The LMK04828 is programmed at start-up over the control plane using \texttt{casperfpga}, from a register set exported from Texas Instruments' TICS Pro and archived with the firmware. Each tile PLL multiplies the 250\,MHz reference to the 1050\,MHz sample clock.

The XCZU48DR's eight converters are not independent: they are grouped into four RF-ADC tiles, numbered 224--227, with two converters sharing each tile. All four tiles are enabled in the deployed build. Tile 226 is configured as the clock-distribution source, forwarding the 250\,MHz reference to tiles 224, 225, and 227, so that every tile PLL derives its sample clock from a single source and no frequency offset can accumulate between converters. This alone does not fix the phase relationship across tiles, since each tile PLL acquires its own start-up phase; deterministic inter-tile alignment requires multi-tile synchronization, which is not engaged in this build and becomes a requirement for the eight-input design (Section~\ref{sec:scaling}). The question does not arise for the two inputs used here, which are the two channels of tile 224 and therefore share one sample clock exactly. A common sample clock is a prerequisite for meaningful cross-correlation: without it the relative phase between the two inputs would drift and the interferometric fringe would decohere. The two routed converters carry East-X polarization (\texttt{auto0}) and West-X polarization (\texttt{auto1}), corresponding to the East and West dishes and the X linear feed polarization in the convention of O'Connor et al.\cite{oconnor2020}.

\subsection{Channelizer}
\label{sec:channelizer}

Each digitized input is channelized by an independent F-engine, following standard practice for PFB spectrometers in radio astronomy \cite{price2018}. Samples are first reordered to present the four-sample-per-clock stream in the ordering expected by the filter bank. Channelization uses a 4-tap PFB
(\texttt{pfb\_fir\_real}\footnote{\url{https://casper-toolflow.readthedocs.io/en/latest/src/blockdocs/Pfb_fir_real.html}}: \texttt{PFBSize}=12, i.e.\ $2^{12}=4096$ points, four simultaneous inputs, unbiased even-value rounding) feeding a real-input FFT of length $N_{\rm FFT} = 4096$ (\texttt{fft\_wideband\_real}\footnote{\url{https://casper-toolflow.readthedocs.io/en/latest/src/blockdocs/Fft_wideband_real.html}}: \texttt{FFTSize}=12, i.e.\ a $2^{12}$-point real transform, four simultaneous inputs, 18-bit twiddle coefficients, 18-bit input with binary point 17, unbiased rounding, wrap-on-overflow). Internal pipeline latencies are configured as (add, multiply, BRAM, convert)$=(1,2,2,1)$ for the filter bank and $(1,3,3,3)$ for the FFT.

The real-input FFT evaluates a single 4096-point real transform using an internal 2048-point complex FFT engine. Consecutive real time samples are packed as the real and imaginary parts of a length-2048 complex sequence, and the block's post-processing stage unpacks the resulting complex spectrum, via the conjugate-symmetry relations of a real-valued input, onto two interleaved output streams, conventionally
labeled ``even'' and ``odd'', each delivering one complex sample per fabric clock over 1024 clocks per spectrum. Recombining them as
\begin{equation}
  X[2k] = X_{\rm even}[k], \qquad X[2k+1] = X_{\rm odd}[k], \qquad k=0,\ldots,1023
\end{equation}
reconstructs $N_{\rm chan}=2048$ channels of width
\begin{equation}
  \Delta\nu = \frac{f_s}{N_{\rm FFT}} = 256.35\,\mathrm{kHz},
\end{equation}
corresponding to $54.1\,\kms$ at the rest frequency of the 21\,cm line. Of the $N_{\rm FFT}/2+1=2049$ unique bins of the real 4096-point transform, the block outputs bins $k=0,\ldots,2047$ (DC through channel 2047) and drops the Nyquist bin ($k=2048$, baseband $f_s/2$). Because Nyquist zone 3 is odd, the channel-to-RF mapping is ascending and non-inverted: $f_{\rm RF}[k] = f_s + k\,\Delta\nu$ for $k=0,\ldots,2047$, i.e.\ 1050.000--1574.744\,MHz. The spectrum repetition period is $N_{\rm FFT}/f_s = 3.901\,\mu$s, or 1024 fabric clocks, giving $2.563\times10^5$ spectra per second at the raw channelizer output, ahead of accumulation.

\subsection{Correlation and accumulation}
\label{sec:correlator}

For $N$ correlated inputs the correlator forms every unique auto- and cross-product of the per-channel complex spectral samples, where $X_i$ denotes the channelized spectrum of input $i$. Auto-products $X_iX_i^*$ are real and non-negative; cross-products $X_iX_j^*$ for $i\neq j$ are complex. Writing $P(N)$ for the number of statistically independent real-valued quantities per channel,
\begin{equation}
  P(N) = \underbrace{N}_{\rm auto} + 2\underbrace{\binom{N}{2}}_{\rm cross} = N^2,
  \label{eq:pn}
\end{equation}
since each of the $N$ auto-products contributes one real quantity and each of the $\binom{N}{2}$ cross-products two. The present two-input instrument therefore produces $P(2)=4$ real quantities per channel: the East and West autocorrelations and the real and imaginary parts of the East$\times$West cross-correlation.

The correlator is organized as four subsystems operating on the even and odd halves of the spectrum in parallel: \texttt{auto\_corr\_0} and \texttt{cross\_corr\_0} consume the even streams of both inputs, and \texttt{auto\_corr\_1} and \texttt{cross\_corr\_1} the odd streams. Each subsystem contains its own power detectors, complex multiplier where applicable, and vector accumulators, and emits two of the eight 64-bit accumulated fields. The complex cross-product is formed by a conjugate complex multiplier immediately upstream of its accumulators; because the on-chip product format is signed while the transport format is an
unsigned bit-reinterpretation adopted for packetization efficiency, the sign must be restored explicitly in host software before the cross-correlation arrays are interpreted.

Accumulation is performed by BRAM-based vector accumulators, each of depth $N_{\rm chan}/2 = 1024$ entries of 64 bits -- one entry per channel index of the even or odd half-spectrum that accumulator serves. Each is a recirculating adder closed through a BRAM delay line whose length matches this vector length, rather than an address-counted memory: a partial sum re-enters the adder exactly one spectrum period after it was written, so the addressing is supplied implicitly by the channelizer's periodic valid-data timing and no address generator is required. 
All eight accumulator instances use the same unsigned fixed-point format, \texttt{UFix\_64\_34} (64 bits wide with 34 fractional bits), so a stored integer $n$ represents $n\times2^{-34}$. For the two auto-products this is the natural format, since they are non-negative by construction. The real and imaginary cross-products are signed, but the unsigned accumulator is nevertheless bit-exact for them: two's-complement addition is identical to unsigned addition modulo $2^{64}$, so the accumulated bit pattern is correct and only its interpretation differs. The signed reading is imposed in host software rather than in firmware (Section~\ref{sec:capture}), which is why sign restoration is required before the cross-correlation arrays are used.

Separately from the streaming path described below, each correlator subsystem also writes its accumulated half-spectrum into a software-readable on-chip memory, one 1024-entry $\times$ 64-bit block per product, addressed by a channel counter cleared at each integration boundary. These memories are read over the control plane and are not part of the 100\,GbE data path; they hold the same accumulated data in addressable form and serve only as a diagnostic path, which the eight-input design of Section~\ref{sec:scaling} removes to free BRAM.

\subsection{Timing and synchronization}
\label{sec:timing}

All logic runs synchronously at $f_{\rm clk}=262.5$\,MHz. Timing is organized as three nested levels, each built on the one above it: a master pulse that sets the fundamental beat, a per-spectrum marker derived from it, and a per-integration boundary derived from that.

\textbf{Master sync:} The instrument needs a single recurring ``start'' pulse that every stage can align to. A counter runs continuously at the clock rate, and each time it reaches a set period it emits a brief pulse and starts over, so the pulse repeats at a fixed, known interval. That period is software-selectable: the
\texttt{sync\_period\_sel} register chooses either a built-in default of $2^{25}-2$ clocks (127.8\,ms) or a custom value in the \texttt{sync\_period\_var} register. Every later stage keys off this one pulse: it enters the filter bank, travels through the channelizer alongside the data, and reappears at the output as \texttt{sync\_out}, which flags the first frequency channel of each spectrum. Because all timing derives from this single reference, the data and its position within a spectrum stay aligned throughout the chain.

\textbf{Spectrum boundary:} The channelizer takes a fixed number of cycles to turn input samples into a finished spectrum, and the downstream correlator stages add a few more. To mark where each spectrum actually emerges from the correlator -- rather than where it entered the channelizer -- the \texttt{sync\_out} pulse is delayed by a matching nine cycles before it is used. This delay is set so that the pulse arrives exactly as the first channel of the corresponding spectrum reaches the accumulators, so each integration starts on a true spectrum boundary.

\textbf{Integration boundary:} An integration is the sum of many consecutive spectra; a dedicated counter decides when one sum ends and the next begins. It counts elapsed cycles while spectra flow and compares that count against the target length, set by the software register \texttt{acc\_len}. Because \texttt{acc\_len} is expressed as a number of spectra while the counter runs in clock cycles, it is scaled up by the 1024 cycles each spectrum occupies before the comparison. When the target is reached, the counter resets to begin the next integration and issues the boundary signal \texttt{new\_acc}; the master pulse also resets it, so the two never drift apart. The integration period is therefore
\begin{equation}
  T_{\rm acc} \simeq \frac{\texttt{acc\_len}\times N_{\rm FFT}/4}{f_{\rm clk}},
\end{equation}
so that \texttt{acc\_len}$=256352$ gives a 1\,s integration and \texttt{acc\_len}$=25632$ gives 100\,ms -- the latter the exact setting used for the 100\,ms run analyzed in Section~\ref{sec:onsky}. At each boundary \texttt{new\_acc} clears the accumulators for the next sum, advances the 64-bit accumulation counter that labels the outgoing spectrum in its packet header, and resets the packet index to zero, so every packet of a given spectrum carries the same accumulation label and a packet index counting up from zero.

\subsection{Visibility packetization}
\label{sec:packetization}

The eight accumulated 64-bit fields are concatenated per channel index by the visibility-packer subsystem into a single 512-bit visibility word: each of the four products contributes two fields, one for the even and
one for the odd half of the spectrum (see Table~\ref{tab:word}). One visibility word is produced per channel index, so a complete spectrum comprises 1024 words, or 64\,KiB. Words are written into a 2048-entry, 512-bit-wide FIFO as they emerge from the accumulators.

\begin{table}[h]
\centering
\caption{512-bit visibility word field map, most significant field first. Cross-correlation fields are transported unsigned and read as signed two's-complement in host software (Section~\ref{sec:correlator}).}
\label{tab:word}
\begin{tabular}{lll}
\toprule
Bits & Field & Channel half \\
\midrule
511:448 & cross imaginary (signed) & odd \\
447:384 & cross real (signed) & odd \\
383:320 & West-X auto, \texttt{auto1} (unsigned) & odd \\
319:256 & West-X auto, \texttt{auto1} (unsigned) & even \\
255:192 & cross imaginary (signed) & even \\
191:128 & cross real (signed) & even \\
127:64  & East-X auto, \texttt{auto0} (unsigned) & odd \\
63:0    & East-X auto, \texttt{auto0} (unsigned) & even \\
\bottomrule
\end{tabular}
\end{table}

Each spectrum is transported as eight UDP datagrams, each carrying one 64\,B header word followed by 128 payload words, so the UDP payload is
\begin{equation*}
  (1+128)\times64\,\mathrm{B} = 8256\,\mathrm{B},
\end{equation*}
transported with no return path (Section~\ref{sec:platform}); with UDP/IP/Ethernet overhead this occupies 8322\,B on the wire (Section~\ref{sec:pacing}), requiring a capture-interface maximum transmission unit (MTU) of at least 8300\,B. The transmit endpoint is configured in firmware rather than at run time: the fabric interface carries a fixed MAC and IP address, and frames are addressed to a fixed destination IP and UDP
port, with all 64 byte lanes of the 512-bit word enabled. Jumbo frames are enabled on the core.

Packet $p$ carries channel indices $128p$ through $128p+127$ for $p=0,\ldots,7$. The header word is a 512-bit big-endian value containing a 32-bit synchronization constant \texttt{0xDEADBEEF}, a 64-bit accumulation counter, an 8-bit packet-index field, an 8-bit total-packet count fixed at 8, and 400 reserved bits set to zero (for future use). In the present design only the low 3 bits of the packet-index field are driven, since eight packets per spectrum is the full range required; Section~\ref{sec:scaling} returns to this at higher
input counts. The synchronization constant permits unambiguous frame identification in a raw capture; the accumulation counter labels the integration and is common to all eight packets of a spectrum; the packet index and total count permit reassembly and loss detection without reference to arrival order.

\subsection{Frame state machine and flow control}
\label{sec:pacing}

\begin{figure}[htbp]
\centering
\includegraphics[width=0.8\linewidth]{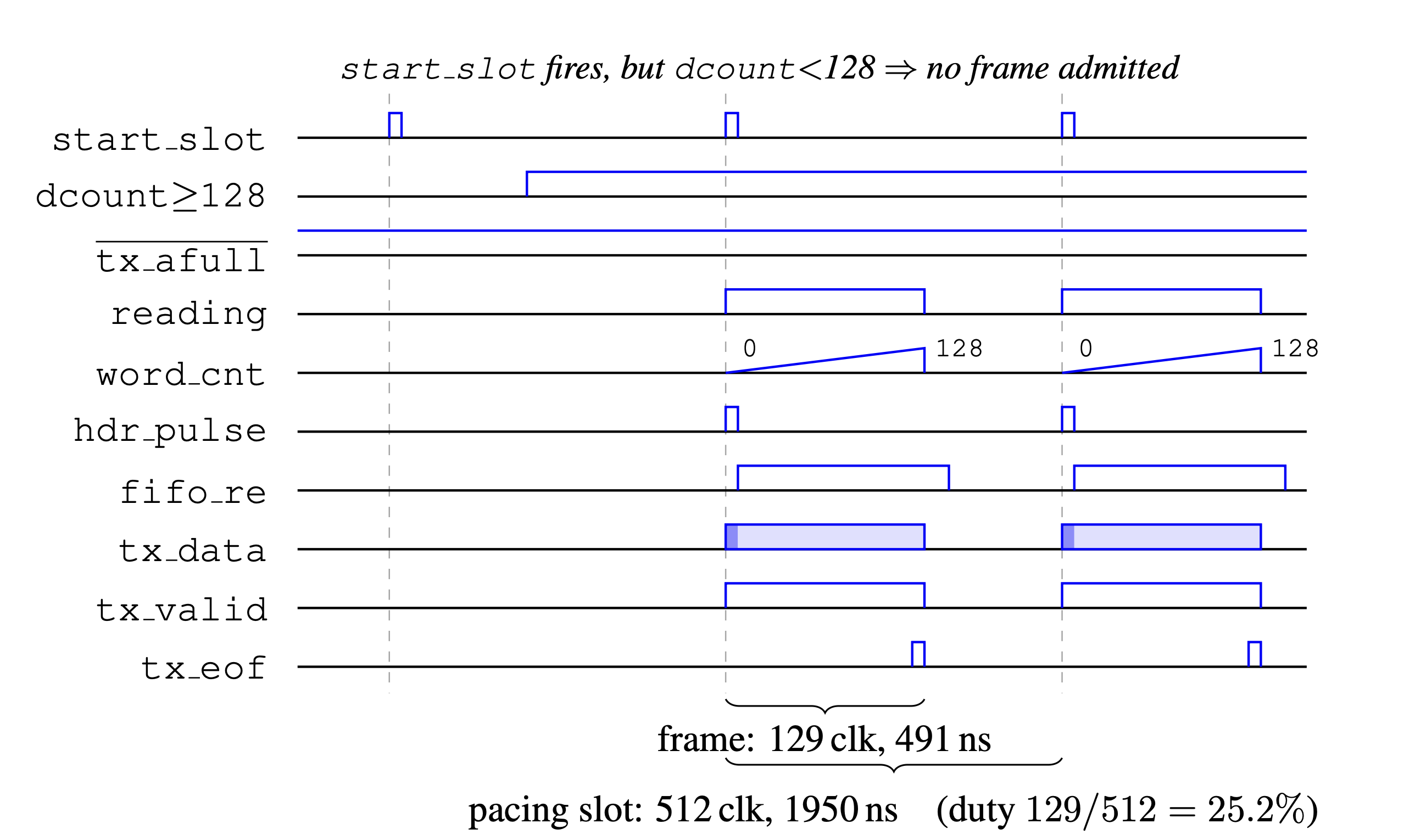}
\caption{Timing of the visibility packetizer (horizontal axis not to scale; asserted levels in blue). A frame start is admitted only when the FIFO holds a complete frame, the 9-bit pacing counter wraps, and the transmit core is not almost-full. The diagram shows the start of one spectrum dump: \texttt{dcount}$\geq$128 rises once the accumulators have written 128 words and falls again after the eighth and last frame of the dump begins, staying low, with the FIFO empty, until the next dump one integration later. Neither readiness term limits streaming in practice: \texttt{dcount}$\geq$128 is low only when there is no complete frame to send, and $\overline{\texttt{tx\_afull}}$ is high throughout, since the 25.2\% duty cycle never brings the core close to full. The pacing slot is therefore the only condition that actually gates frame starts in steady state. Each frame occupies 129 consecutive valid cycles -- one header word plus 128 payload words -- and drains onto the wire in 175 clocks, leaving the link idle for the remainder of the 512-clock pacing slot. \texttt{word\_cnt} returns to zero when \texttt{reading} clears, and \texttt{tx\_eof} is asserted on the final payload word rather than on an idle cycle. The resulting 25.2\% duty cycle provides a 2.93$\times$ margin over the 175-clock line-rate drain time of each frame.}
\label{fig:packetizer}
\end{figure}

Rather than draining the FIFO continuously, packet framing is implemented by an explicit read-side finite-state machine. A \texttt{reading} register holds the frame-active state; a 129-state word counter, enabled by \texttt{reading} and reset when \texttt{reading} is low or on global reset, tracks position within the frame as shown in Figure~\ref{fig:packetizer} (the signals are summarized in Table~\ref{tab:pktsignals}).

\begin{table}[htbp]
\centering
\caption{Signals in the visibility-packetizer timing diagram (Figure~\ref{fig:packetizer}). A frame start is admitted only when $\texttt{start\_slot}\wedge(\texttt{dcount}\!\geq\!128)\wedge \overline{\texttt{tx\_afull}}$; once admitted, \texttt{reading} holds the frame active for its full 129 cycles.}
\label{tab:pktsignals}
\begin{tabular}{@{}lp{0.72\linewidth}@{}}
\toprule
Signal & Meaning and role in the frame handshake \\
\midrule
\texttt{start\_slot}
  & One-cycle pulse at the start of every 512-clock pacing slot (the
    9-bit pacer counter wrapping to zero). The only instant at which a
    new frame may begin; ANDed with the two readiness terms below. \\[2pt]
\texttt{dcount}$\geq$\texttt{128}
  & High when FIFO occupancy is at least one full frame (128 payload
    words), so a frame can be emitted without underflow -- the
    \emph{data-available} term. High from the 128th word of each spectrum dump until the
    dump's last frame begins; low between dumps, when the FIFO is empty. \\[2pt]
$\overline{\texttt{tx\_afull}}$
  & The \emph{core-ready} term, formed by registering and inverting the 100\,GbE core's almost-full flag; high means the core can accept a frame. The
    light load keeps it continuously high in steady state. \\[2pt]
\texttt{reading}
  & Frame-active register. Set when a frame start is admitted, held for
    the whole frame by word-counter feedback, cleared when
    \texttt{word\_cnt} reaches 128. Gates \texttt{fifo\_re} and
    \texttt{tx\_valid}. \\[2pt]
\texttt{word\_cnt}
  & 129-state counter (0--128), enabled by \texttt{reading}, tracking
    position within the frame: 0 is the header word, 1--128 the payload
    words. Resets when \texttt{reading} clears. \\[2pt]
\texttt{hdr\_pulse}
  & One-cycle pulse at \texttt{word\_cnt}$=0$. Multiplexes the 64\,B
    header onto \texttt{tx\_data} and masks \texttt{fifo\_re}, so the
    header consumes no payload word. \\[2pt]
\texttt{fifo\_re}
  & FIFO read enable. Asserted throughout \texttt{reading} except at
    \texttt{word\_cnt}$=0$ (masked by \texttt{hdr\_pulse}), so exactly
    128 payload words are popped per frame. \\[2pt]
\texttt{tx\_data}
  & 512-bit word presented to the core: the header on
    \texttt{word\_cnt}$=0$ (dark), payload words on 1--128 (light). \\[2pt]
\texttt{tx\_valid}
  & Qualifies \texttt{tx\_data}; high for all 129 words of the frame,
    low during the inter-frame gap. \\[2pt]
\texttt{tx\_eof}
  & End-of-frame strobe, asserted with the final payload word
    (\texttt{word\_cnt}$=128$). Clears \texttt{reading}; the link then
    idles for the rest of the pacing slot, forming the inter-frame
    gap. \\
\bottomrule
\end{tabular}
\end{table}

A frame may begin only when the logical AND of three conditions is satisfied: (1) \emph{data available} -- FIFO occupancy is at least 128 words, so a complete frame can be emitted without underflow; (2) \emph{pacing slot} -- a free-running 9-bit counter equals zero, so frames may begin only on one clock in every 512; (3) \emph{core ready} -- the 100\,GbE core's \texttt{tx\_afull} almost-full flag is registered and inverted to form a core-ready signal, \texttt{ready = \textasciitilde tx\_afull}, and a frame may start only when \texttt{ready} is high (equivalently, \texttt{tx\_afull} is low). Once armed, the state is held for the duration of the frame by feedback through the word counter, and is released when the counter reaches 128. In the deployed build the pacing period is fixed at 512 clocks by the width of this counter; for the scaling study of Section~\ref{sec:scaling} we additionally built a variant in which the counter is widened by one bit and cleared by its own comparison against a software register, making the period settable at run time from the control plane without a rebuild.

Within a frame, the header word is multiplexed onto the data bus on word count 0, gated by both the frame-start pulse and the \texttt{reading} state; on that same cycle the FIFO read enable is masked, so the header slot consumes no payload word and the number of payload words per frame is structurally guaranteed to be 128 rather than timing-dependent. The end-of-frame strobe is generated when the word counter reaches 128 and is delayed through two registers so that it is coincident with the final payload word on the wire.

\subsubsection*{Why pacing is required}
The pacing term is not a performance optimization but a correctness requirement, and the reason is not evident from the transmit core's interface. The visibility bus is 512 bits wide and can present one word
per clock, giving an instantaneous offered rate of
\begin{equation}
  R_{\rm bus} = 512\,\mathrm{bit}\times262.5\,\mathrm{MHz} = 134.4\,\mathrm{Gb\,s^{-1}},
  \label{eq:rbus}
\end{equation}
which exceeds the 100\,Gb\,s$^{-1}$ line rate by 34\%. In an initial implementation in which frames were emitted as soon as data were available, with no enforced inter-frame gap, we observed systematic corruption in captured traffic: the first frame following an idle period was always well formed, but subsequent back-to-back frames were damaged, with duplicated header words occupying two consecutive word slots, insertion of null words, and outright loss of frames from the sequence. Introducing a single-clock idle gap reduced but did not eliminate the corruption. The pattern -- correct after idle, corrupt when contiguous, with duplication and insertion rather than truncation -- indicates a frame-boundary handshake race inside the transmit core rather than a byte-level buffer overflow. We note, however, that both configurations tested lie \emph{below} the 175-clock line-rate floor of Eq.~(\ref{eq:tdrain}), offering 134.4--135.5\,Gb\,s$^{-1}$ against a $100\,\mathrm{Gb\,s^{-1}}$ link, so sustained oversubscription is not excluded as an alternative
explanation. Distinguishing the two requires a pacing period between 175 and 512 clocks, which we have not tested; Section~\ref{sec:scaling} describes a run-time-settable pacer that makes such a sweep a single-bitstream experiment.

The resolution is to pace frame starts so that each frame has drained before the next begins. A frame of 8256\,B payload plus 66\,B of UDP, IP, Ethernet and inter-packet-gap overhead occupies 8322\,B on the wire,
requiring
\begin{equation}
  T_{\rm drain} = \frac{8322\times8}{10^{11}} = 666\,\mathrm{ns} = 175\,\mathrm{clk}
  \label{eq:tdrain}
\end{equation}
at line rate. Admitting a frame start only once per 512 clocks (1950\,ns) therefore provides a margin of 2.93$\times$ and limits the input-bus valid-cycle duty cycle to $129/512=25.2\%$ of the $134.4\,\mathrm{Gb\,s^{-1}}$ bus rate of Eq.~(\ref{eq:rbus}), corresponding to an average of $\approx33.9\,\mathrm{Gb\,s^{-1}}$ of UDP payload over each pacing slot -- about 34\% of the $100\,\mathrm{Gb\,s^{-1}}$ MAC line rate, averaged over the pacing slot rather than sustained during the active burst. The \texttt{tx\_afull} interlock provides an independent backstop against residual congestion. A complete spectrum is read out in $8\times512=4096$ clocks, or $15.6\,\mu$s, a negligible fraction of any accumulation period.

\subsection{Verification}
\label{sec:verification}
Because packetizer defects manifest as intermittent wire-format errors rather than as build failures, the design was verified against a cycle-accurate acceptance test before deployment. A counter-based fingerprint pattern was injected in place of the visibility payload, with a delay stage mimicking the accumulator read latency so that the test exercised the same timing alignment as the real data path. The following invariants were checked in simulation: the first payload word of every frame is the expected sequence start; the last payload word before the final end-of-frame of a spectrum is the expected sequence end; exactly 129 valid cycles occur per frame, with one idle gap between consecutive frames; the header occupies the first word of every frame, exactly once, and is a single cycle wide; mid-spectrum frame seams are contiguous, with no payload word skipped or repeated across the header slot; the end-of-frame strobe is coincident with the final payload word rather than with an idle cycle; and FIFO occupancy returns to exactly zero at the end of every integration.

The last of these is a strong global invariant: any imbalance between words written by the accumulator and words consumed by the packetizer accumulates monotonically and is immediately visible as a drift in FIFO
occupancy from one integration to the next. Its satisfaction, together with the wire-level packet integrity results reported in Section~\ref{sec:capture}, is the basis of the zero-dropped-accumulation performance of the deployed system.

\begin{table}[h]
\centering
\caption{Firmware parameters of the deployed model.}
\label{tab:firmware}
\begin{tabular}{lll}
\toprule
Quantity & Expression & Value \\
\midrule
Sample rate & $f_s$ & 1050\,MHz \\
Samples per fabric clock & -- & 4 \\
Fabric clock & $f_s/4$ & 262.5\,MHz (3.810\,ns) \\
Nyquist zone & -- & third (1050--1575\,MHz) \\
PFB taps & -- & 4 \\
FFT length & $N_{\rm FFT}$ & 4096 (real) \\
Sky channels & $N_{\rm chan}$ & 2048 \\
Channel width & $f_s/N_{\rm FFT}$ & 256.35\,kHz \\
Spectrum period & $N_{\rm FFT}/f_s$ & 3.901\,$\mu$s (1024\,clk) \\
Products per channel & $P(2)$ & 4 \\
Accumulator format (all eight) & -- & \texttt{UFix\_64\_34} \\
Cross-product host interpretation & -- & signed two's complement \\
Accumulator depth & $N_{\rm chan}/2$ & 1024 \\
Master sync period (default) & $2^{25}-2$ & 127.8\,ms \\
Visibility word & $8\times64$\,bit & 512\,bit (64\,B) \\
FIFO & -- & $2048\times512$\,bit \\
Frame length & $1+128$ words & 129\,clk (491\,ns) \\
Pacing period & $2^9$ & 512\,clk (1950\,ns) \\
Transmit duty cycle & $129/512$ & 25.2\% \\
Offered bus rate & Eq.~(\ref{eq:rbus}) & 134.4\,Gb\,s$^{-1}$ \\
Frame drain at line rate & Eq.~(\ref{eq:tdrain}) & 666\,ns (175\,clk) \\
Pacing margin & $512/175$ & 2.93$\times$ \\
Spectrum readout & $8\times512$\,clk & 15.6\,$\mu$s \\
UDP payload per packet & $129\times64$\,B & 8256\,B \\
Packets per spectrum & -- & 8 \\
\bottomrule
\end{tabular}
\end{table}

The deployed two-input design occupies a small fraction of the XCZU48DR (Table~\ref{tab:util}): 6.9\% of CLB LUTs, 5.9\% of registers, 3.4\% of DSP48E2 slices, and 15.1\% of BRAM. 
These comfortable margins indicate the two-input instrument is far from any device limit and motivate the scaling analysis of Section~\ref{sec:scaling}.

\begin{table}[htbp]
\centering
\caption{Post-implementation resource utilization of the deployed
two-input design.}
\label{tab:util}
\begin{tabular}{lrrr p{0.42\linewidth}}
\toprule
Resource & Used & Available & Util.\ & Remark \\
\midrule
CLB LUTs        & 29\,452 & 425\,280 & 6.9\%  & Configurable Logic Block Look-Up Tables. General-purpose logic implementing the control and glue logic. \\
CLB registers   & 50\,322 & 850\,560 & 5.9\%  & Pipeline flip-flops holding intermediate values between clock cycles. \\
DSP48E2         & 144     & 4\,272   & 3.4\%  & Hardened multiply--accumulate slices used by the FFT, PFB, and correlator. \\
BRAM tiles & 162.5   & 1\,080   & 15.1\% & On-chip memory for PFB coefficients, FFT buffers, vector accumulators, and the packet FIFO. \\
RF-ADC tiles    & 4       & 4        & 100\%  & Direct-sampling data-converter tiles; all four enabled, two converters routed to the DSP chain. \\
CMAC (100\,GbE) & 1       & 2        & 50\%   & Hardened 100\,GbE Ethernet MAC driving the visibility output link. \\
GTY channels    & 4       & 16       & 25\%   & High-speed serial transceiver lanes carrying the 100\,GbE link. \\
\bottomrule
\end{tabular}
\end{table}

\section{Data Capture and Streaming Validation}
\label{sec:capture}
Each 64\,B word arrives byte-reversed on the wire; the capture software detects the byte order automatically, by locating the synchronization constant \texttt{0xDEADBEEF} under each candidate ordering. A Python capture script listens on the UDP port, groups received packets by the header's accumulation counter and packet index into an in-memory table, and assembles a spectrum only once all eight packets for a given accumulation counter are present; a late or duplicate packet for an already-complete accumulation is simply never consulted, and incomplete accumulation groups older than two accumulation counts are discarded unassembled rather than written. Because assembly requires the complete set of eight packets, no partial or corrupted spectrum can be written to disk: an accumulation either appears whole or does not appear at all, and the script separately counts malformed packets (wrong length, bad sync, or invalid packet index/count fields) and un-assembled (incomplete) accumulation groups. The script writes time-chunked HDF5 files containing \texttt{auto0}, \texttt{auto1} (East/West autocorrelations), \texttt{cross\_real}/\texttt{cross\_imag} (sign-restored cross-correlation), the monotonic \texttt{acc\_count}, the host-arrival time, as Unix seconds (\texttt{timestamp}) and as an ISO-8601 UTC string (\texttt{timestamp\_utc}), and a stored frequency axis. Raw correlator counts are converted to field values using the $\texttt{UFix\_64\_34}$ scaling of Section~\ref{sec:correlator}; because this conversion does not divide by \texttt{acc\_len}, stored and plotted powers are accumulated \emph{sums} of \texttt{acc\_len} raw spectra rather than per-spectrum averages, so absolute power levels are not directly comparable between the 500\,ms and 100\,ms campaigns (which used different \texttt{acc\_len}) without an explicit normalization.

Across both runs, the accumulation counter is unbroken: 317,817 (run\#1) and 641,101 (run\#2) accumulations, 958,918 in total, with the counter span in each run matching the spectrum count exactly -- zero gaps and
zero duplicates -- corresponding to 2,542,536 and 5,128,808 packets (7,671,344 total) and 19.71 and 39.75\,GiB (59.46\,GiB total) of visibility traffic on the wire (payload plus UDP/IP/Ethernet overhead, Section~\ref{sec:pacing}). Given the assembly logic above, this is a meaningful zero-loss statistic: any dropped, duplicated, or malformed packet would either prevent an accumulation from ever completing (leaving a gap in \texttt{acc\_count}) or be silently absorbed without altering which packets contribute (duplicates), never producing a wrong-but-complete spectrum -- and it is consistent with the pre-deployment verification described in Section~\ref{sec:verification}.

\subsection{End-to-end format validation}
\label{sec:cauchyschwarz}
The on-sky results of Section~\ref{sec:onsky} confirm that the deployed system recovers physically sensible signals, but they do not by themselves rule out an internally self-consistent misinterpretation of Table~\ref{tab:word} -- for instance, a swap of the two auto fields -- which could still pass those checks. The Cauchy--Schwarz inequality gives an independent, sky-content-independent test. For each frequency channel of each accumulation, form the normalized cross-correlation coefficient
\begin{equation}
  r = \frac{|V_{EW}|}{\sqrt{P_E P_W}},
  \qquad
  |V_{EW}| = \sqrt{(\mathrm{Re}\,V_{EW})^2 + (\mathrm{Im}\,V_{EW})^2},
  \label{eq:cauchy}
\end{equation}
where $P_E$ and $P_W$ are the East and West autocorrelation powers (\texttt{auto0}, \texttt{auto1}) and $\mathrm{Re}\,V_{EW}$, $\mathrm{Im}\,V_{EW}$ are the real and imaginary parts of the East$\times$West cross-correlation (stored as \texttt{cross\_real} and \texttt{cross\_imag}), all taken from the same channel of the same accumulation and placed on the common $2^{-34}$ scale. Cauchy--Schwarz requires $0\le r\le1$ identically, regardless of calibration or sky content. The bound is violated if any of the four 64-bit fields is mis-assigned within the 512-bit word, if byte-order detection fails, if the even and odd channel halves are crossed between products, or if the unsigned and signed accumulator formats are not placed on that common
scale before use.

We form $r$ for every channel of every accumulation across both campaigns: $1{,}963{,}864{,}064$ channel-samples in total, with zero violations of the bound. The overall maximum, $r=0.99915$, occurs in the
DC channel (1050.000\,MHz), which lies below the analog passband; within the 1100--1550\,MHz passband the maximum is $r=0.99913$ (at 1114.09\,MHz), and more than 540 of the 1755 passband channels exceed $r=0.9$ at some point in each campaign. Because the bound is approached within the passband, across many channels, the test is exercised near its limit rather than being vacuously satisfied by a low-amplitude cross
term. This validates the visibility word map, byte-order detection, and the placement of all four transported fields on a common scale, but by construction depends only on cross-correlation \emph{magnitude}: it does not constrain the sign convention used to interpret \texttt{cross\_real}/\texttt{cross\_imag} (Section~\ref{sec:further}).

\subsection{Noise performance}
\label{sec:radiometer}

As a further check that the correlator behaves as a linear, thermal-noise-limited system, we measured the per-channel noise from successive differences of the autocorrelations in RFI-free channels. The fractional rms
is $0.90$--$0.91$ of the naive radiometer prediction $1/\sqrt{\Delta\nu\,T_{\rm acc}}$ in run\#2, and centered around $0.91$--$0.92$ (with greater file-to-file scatter, 16th--84th percentile range $0.89$--$0.96$) in run\#1. A ratio below unity is expected for a polyphase channelizer, whose channel responses overlap; the
corresponding equivalent noise bandwidth is $1.18$--$1.24\,\Delta\nu$ (303--318\,kHz) against the 256.35\,kHz channel spacing, consistent between campaigns. The implied adjacent-channel noise correlation is more variable than a single figure conveys -- pooled medians of $\rho_1\approx0.03$ (run\#2) and $\approx0.06$ (run\#1), with individual files ranging up to $\rho_1\sim0.15$ -- and is used only qualitatively in Section~\ref{sec:hi}.

Table~\ref{tab:measured} collects these measured performance figures alongside the design parameters of Table~\ref{tab:firmware}, as a companion record of what was actually observed rather than only specified.

\begin{table}[h]
\centering
\caption{Measured performance, both campaigns combined unless noted.}
\label{tab:measured}
\begin{tabular}{ll}
\toprule
Quantity & Measured value \\
\midrule
Accumulations delivered / lost & 958,918 / 0 \\
Packets delivered / lost & 7,671,344 / 0 \\
Visibility traffic on the wire & 59.46\,GiB \\
$T_{\rm acc}$, run\#1 / run\#2 & 499.9945\,ms / 99.9964\,ms \\
Accumulation interval jitter (host), run\#1 / run\#2 & 37.8 / 14.5\,$\mu$s (robust rms) \\
Cauchy--Schwarz violations & 0 of $1.96\times10^9$ samples \\
Maximum normalized correlation $r$ & 0.99915 (passband: 0.99913) \\
Noise vs.\ radiometer prediction, run\#1 / run\#2 & $\sim$0.91--0.92 / 0.90--0.91 \\
PFB equivalent noise bandwidth & 1.18--1.24\,$\times\Delta\nu$ \\
Adjacent-channel noise correlation $\rho_1$ & $\sim$0.03--0.06 (up to 0.15) \\
\bottomrule
\end{tabular}
\end{table}

\section{Observations}
\label{sec:observations}

Two zenith drift-scan campaigns are used (Table~\ref{tab:datasets}), both at BNL with 2048 channels over 1050--1575\,MHz, differing only in accumulation time and duration.

\begin{table}[h]
\centering
\caption{Data used in this paper.}
\label{tab:datasets}
\begin{tabular}{llllr}
\toprule
Run & Accum. & Span (UTC) & Duration & Accumulations \\
\midrule
run\#1 & 500\,ms & 2026-07-24 23:21 -- 07-26 19:29 & 44.1\,hr & 317,817 \\
run\#2 & 100\,ms & 2026-07-27 19:25 -- 07-28 13:13 & 17.8\,hr & 641,101 \\
\bottomrule
\end{tabular}
\end{table}

Specific subsets: the GNSS check (Section~\ref{sec:gnss}) uses a 10-hour subset of run\#1 (2026-07-25 16:57\,UTC to 2026-07-26 02:57\,UTC). The Galactic \HI{} check (Section~\ref{sec:hi}) uses an earlier, non-overlapping 5-hour subset of run\#1 (2026-07-25 10:57--15:57\,UTC), bracketing the zenith beam's plane crossing at 13:46\,UTC ($l=165.6^\circ$). The \cygA{} check (Section~\ref{sec:cygnusa}) uses run\#1 2026-07-26 01:57--06:57\,UTC (bracketing the 04:36:18\,UTC transit) and run\#2 2026-07-28 01:13--07:13\,UTC (bracketing the 04:28:27\,UTC transit), each providing at least $\sim$2\,hr of baseline on either side of the
transit.

\section{On-Sky Validation}
\label{sec:onsky}

With the backend design validated at the wire-format level (Section~\ref{sec:verification}) and the capture chain validated at the packet level (Section~\ref{sec:capture}), we use three independent celestial signals to confirm that the deployed system produces complete, physically sensible visibilities on real sky data: two kinds of compact source (GNSS-band transits, \cygA{}) checked in cross-correlation, and one extended, diffuse source (Galactic \HI{}) checked in autocorrelation.

\subsection{GNSS transits}
\label{sec:gnss}

GNSS constellations broadcast strong, narrowband signals across several cataloged L-band lines searched within our RF band (GPS L2/L5, GLONASS L3, Galileo E5a/E5b/E6, BeiDou B2b/B3), with band centers taken from
the respective constellation interface control documents. Because the beam is fixed at zenith, satellites enter and leave it through a combination of Earth's rotation and their own orbital motion, each transit appearing as a transient enhancement in the autocorrelations with correlated fringe structure in the
cross-correlation.

Events are found per line as follows. For each line's band-averaged power (mean of the two autocorrelations over the channels within its half-width, Table~\ref{tab:gnss}), a slowly varying local floor is estimated as a 15th-percentile rolling filter over a $\sim$30-minute window, long enough to track gain drift without absorbing a transit; $\sigma$ is the robust median absolute deviation of (power $-$ floor) about its own median. A candidate event is a local maximum, at least 60\,s from any other, with fractional excess over the floor $\geq$50\% and absolute excess $\geq6\sigma$; its FWHM comes from a local Gaussian fit over a $\pm$120\,s window. An event is called \emph{fringe-confirmed} if the cross-correlation real part's standard deviation over the same window, normalized by the mean band power there, exceeds 2\%. This search finds 141 transit-like events across six active lines (listed in Table~\ref{tab:gnss}), 140 of which (99\%) are fringe-confirmed by this criterion -- the expected interferometric signature of a compact source, and evidence that both amplitude and phase are being captured and correlated correctly end-to-end. We report these as GNSS-band transit-like events rather than as identified satellite passes: a separate cross-match against GPS/GLONASS/Galileo/BeiDou two-line-element ephemerides (for beam-mapping purposes, not shown here) positively identified the transiting satellite for 164 of the 507 such events found over run\#1, with zenith distance at the matched time $\leq75^\circ$; the remaining events are not individually ephemeris-confirmed. Adjacent search windows overlap (GLONASS L3 with Galileo E5b/BeiDou B2b, and BeiDou B3 with Galileo E6), so one emitter can be counted in more than one line: merging detections within 20\,s of one another reduces the 141 events to 56 distinct ones. Most of the short events (FWHM 10--25\,s) recur in trains at $\sim$6-minute
intervals, and the fringe rates of those examined imply East--West angular rates of 0.15--0.38$^\circ$\,s$^{-1}$, 50--100 times faster than a satellite in medium Earth orbit; they are consistent with low-Earth-orbit emitters in the GNSS bands rather than with GNSS satellites. They serve this check equally well, since each is a compact source fringing coherently on the East--West baseline. As a coarse, independent check on the
channel-to-RF mapping, all six detected lines' band-averaging windows (Table~\ref{tab:gnss}) are centered on their cataloged frequencies to within the windows' own half-widths; this confirms the mapping at the tens-of-MHz level exercised by the search itself, not at the single-channel precision a dedicated swept-tone test (Section~\ref{sec:further}) would provide.

\begin{table}[h]
\centering
\caption{GNSS L-band transit inventory, 10-hr run\#1 subset. HW is the half-width of the frequency window over which each line's power is averaged (Freq.\ $\pm$ HW). Counts are per line, without merging detections in overlapping windows (see text). FWHM is the full width at half maximum, in time, of a Gaussian fitted to each event's band power over $\pm$120\,s; the mean is over events with a successful fit (4 of 8 for GPS L5, 2 of 5 for GPS L2, 25--35 per line elsewhere), and widths beyond $\sim$240\,s exceed the fit window and are approximate. The GPS L5 and L2 means are dominated by slow medium-Earth-orbit transits, the other lines by short, fast-moving events. GPS L2's single fringe-confirmation failure is not explained.}
\label{tab:gnss}
\small
\begin{tabular}{lrrrrr}
\toprule
Line & Freq.\ (MHz) & HW (MHz) & Events & Mean FWHM (s) & Fringe-confirmed \\
\midrule
GPS L5 / Galileo E5a     & 1176.45 & 12 &  8 & 225.3 &  8/8  \\
GLONASS L3               & 1202.03 &  5 & 38 &  72.6 & 38/38 \\
Galileo E5b / BeiDou B2b & 1207.14 &  6 & 35 &  61.2 & 35/35 \\
GPS L2                   & 1227.60 & 12 &  5 & 398.1 &  4/5  \\
BeiDou B3                & 1268.52 & 10 & 28 &  51.1 & 28/28 \\
Galileo E6               & 1278.75 & 10 & 27 &  68.7 & 27/27 \\
\midrule
\textbf{Total}           &         &    & \textbf{141} & & \textbf{140/141} \\
\bottomrule
\end{tabular}
\end{table}

\begin{figure}[htbp]
\centering
\includegraphics[width=\linewidth]{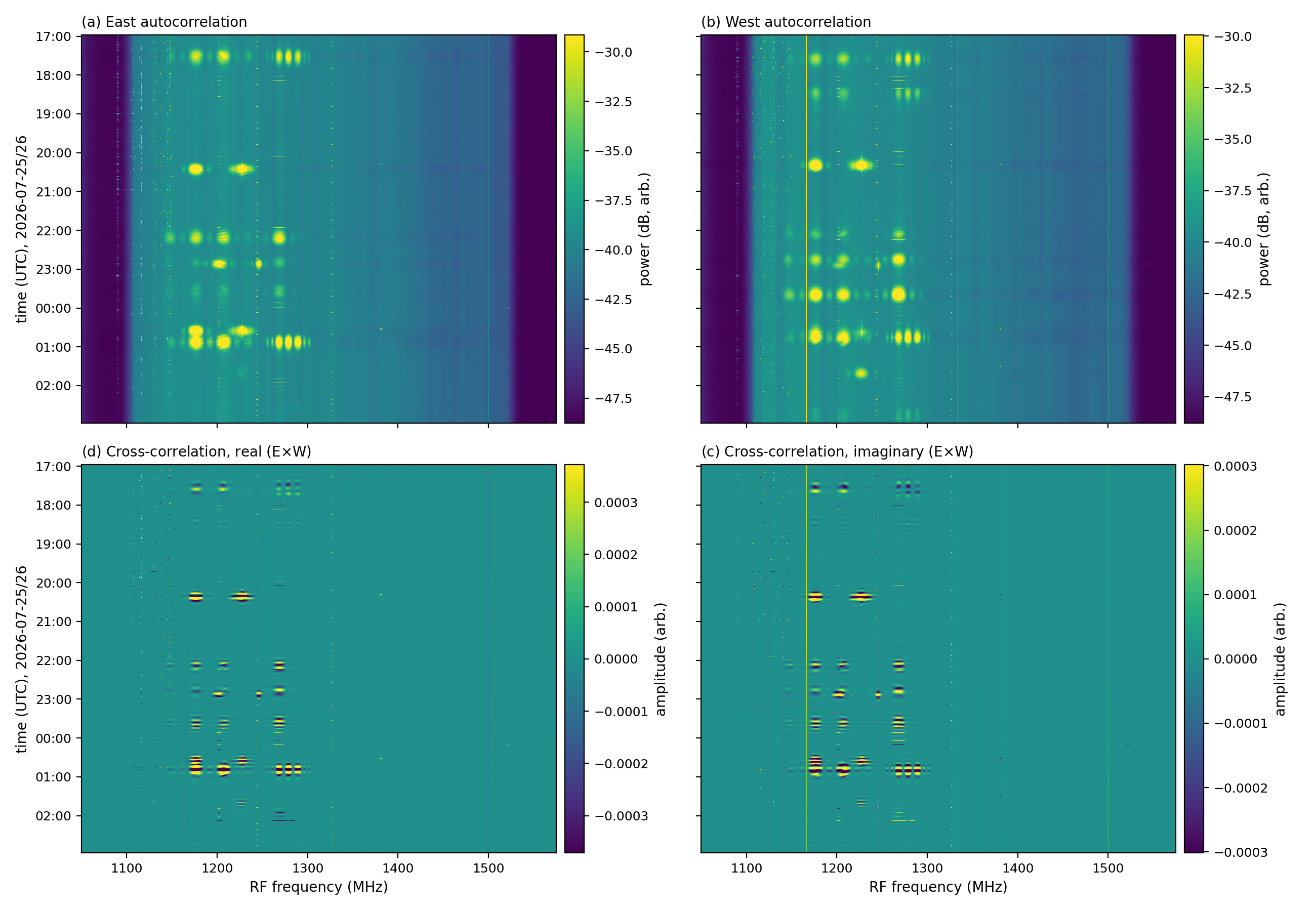}
\caption{The four correlator data products over the 10-hour run\#1 subset (2026-07-25 16:57 to 07-26 02:57\,UTC); time runs downward. (a)~East autocorrelation and (b)~West autocorrelation, in dB (arbitrary reference). (c)~Imaginary and (d)~real parts of the East$\times$West cross-correlation, on a symmetric linear scale. GNSS-band transits appear as repeated bright patches in the autocorrelations and as matching fringe patterns in the cross terms, in the 1150--1300\,MHz range.}
\label{fig:waterfalls}
\end{figure}

Figure~\ref{fig:waterfalls} shows the four correlator data products over this subset: GNSS transits are visible as bright, repeated streaks in the 1150--1300\,MHz range, with matching fringe structure in the cross-correlation.

\begin{figure}[htbp]
\centering
\includegraphics[width=\linewidth]{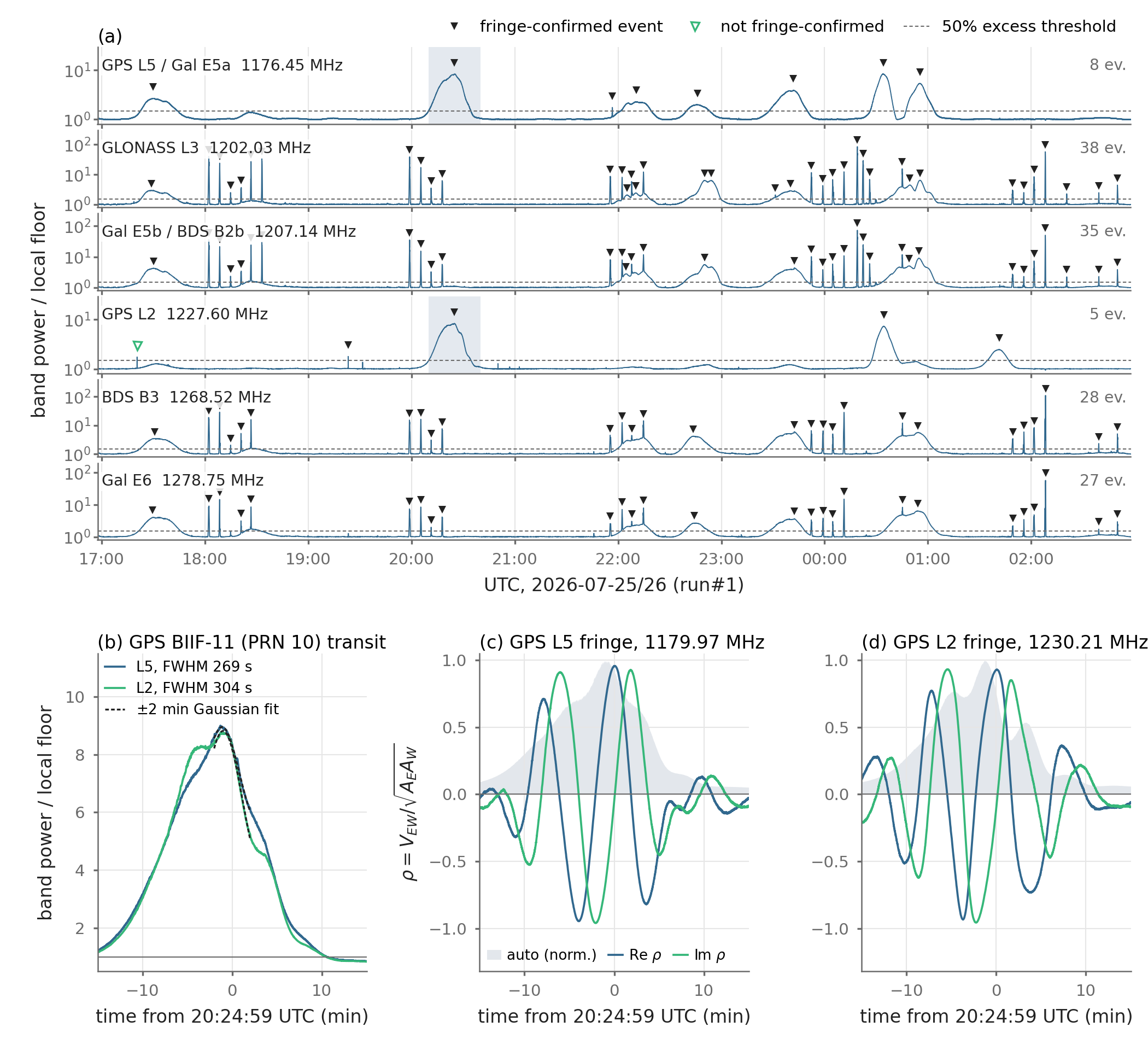}
\caption{GNSS transit detection in the 10-hour run\#1 subset. (a)~Band-averaged autocorrelation power divided by its rolling 15th-percentile floor for the six active lines (Table~\ref{tab:gnss}); triangles mark detected events (filled: fringe-confirmed; open: the one GPS L2 event that fails the fringe criterion), and the dashed line is the 50\% fractional-excess threshold. Shaded: the window expanded below. (b)~Band power for the GPS BIIF-11 (PRN~10) transit in GPS L5 and L2, with the $\pm$120\,s Gaussian fits used for the FWHM in Table~\ref{tab:gnss}. (c,d)~Real and imaginary parts of the normalized cross-correlation $\rho$ in the single channel with the strongest fringe in each line, over the normalized autocorrelation (shaded).}
\label{fig:gnss}
\end{figure}

Figure~\ref{fig:gnss} shows the event search itself. Panel~(a) plots each line's band power relative to its rolling floor over the same subset, with the 141 detected events marked. Panels~(b)--(d) zoom on one ephemeris-identified transit, GPS BIIF-11 (PRN~10), which passes $1.0^\circ$ from zenith at 20:24:59\,UTC and is detected independently in both GPS L5 and GPS L2. In a single channel of each line the normalized cross-correlation $\rho=V_{EW}/\sqrt{P_E P_W}$ oscillates with $|\rho|$ exceeding 0.9 near the peak of the transit envelope, and its real and imaginary parts stay in quadrature throughout. This is the fringe of a compact source crossing the E--W baseline.

\subsection{The \cygA{} transit}
\label{sec:cygnusa}

\cygA{} (ICRS $\alpha=19^{\rm h}59^{\rm m}28.36^{\rm s}$, $\delta=+40^\circ44'02.1''=+40.7339^\circ$) transits BNL's zenith at a minimum zenith distance of $0.06^\circ$. The predicted transit times and minimum zenith distances quoted in this section are computed rigorously: the ICRS position is transformed with \texttt{astropy} \cite{astropy2022} into the topocentric altitude/azimuth frame at each trial time, and the transit is the time of maximum altitude over a dense grid, which correctly incorporates precession and nutation via \texttt{astropy}'s frame transformation. Predicted transits by this method fall at 2026-07-26 04:36:18\,UTC (run\#1) and 2026-07-28 04:28:27\,UTC (run\#2, $\sim$4\,minutes earlier per day, as expected, and two days after the run\#1 transit -- the two campaigns themselves start three days apart, Section~\ref{sec:observations}). In both windows the Sun is $>8^\circ$ below the horizon throughout, the Moon stays $>65^\circ$ off zenith, and no other cataloged bright radio source approaches zenith.

\begin{figure}[htbp]
\centering
\includegraphics[width=\linewidth]{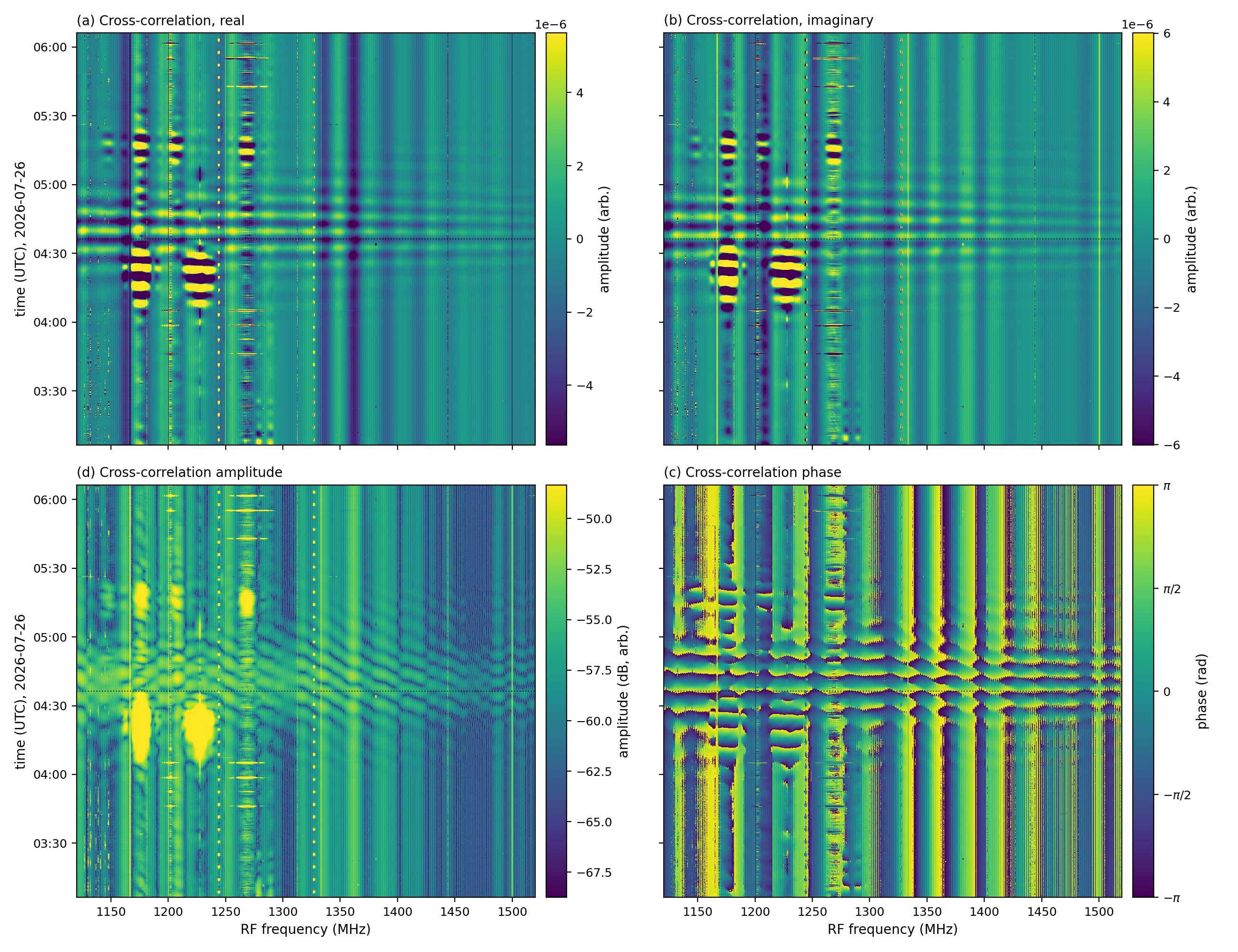}
\caption{Cross-correlation waterfall through the run\#1 \cygA{} transit, $\pm90$\,min around the predicted transit time (04:36:18\,UTC, dotted line), 1120--1520\,MHz with no channels flagged. GNSS passes appear in the 1160--1290\,MHz range (e.g.\ near 04:20 and 05:15\,UTC). (a)~Real and (b)~imaginary parts of the East$\times$West cross-correlation, on linear color scales set from RFI-free channels, so the GNSS passes saturate; (c)~its phase; (d)~its amplitude in dB, with the color range set from RFI-free channels to bring out the \cygA{} fringes; the GNSS passes, up to $\sim$40\,dB brighter, saturate. The fringe rate across frequency passes through zero at the predicted transit time, within the plot's time resolution.}
\label{fig:cygfringes}
\end{figure}

\begin{figure}[htbp]
\centering
\includegraphics[width=\linewidth]{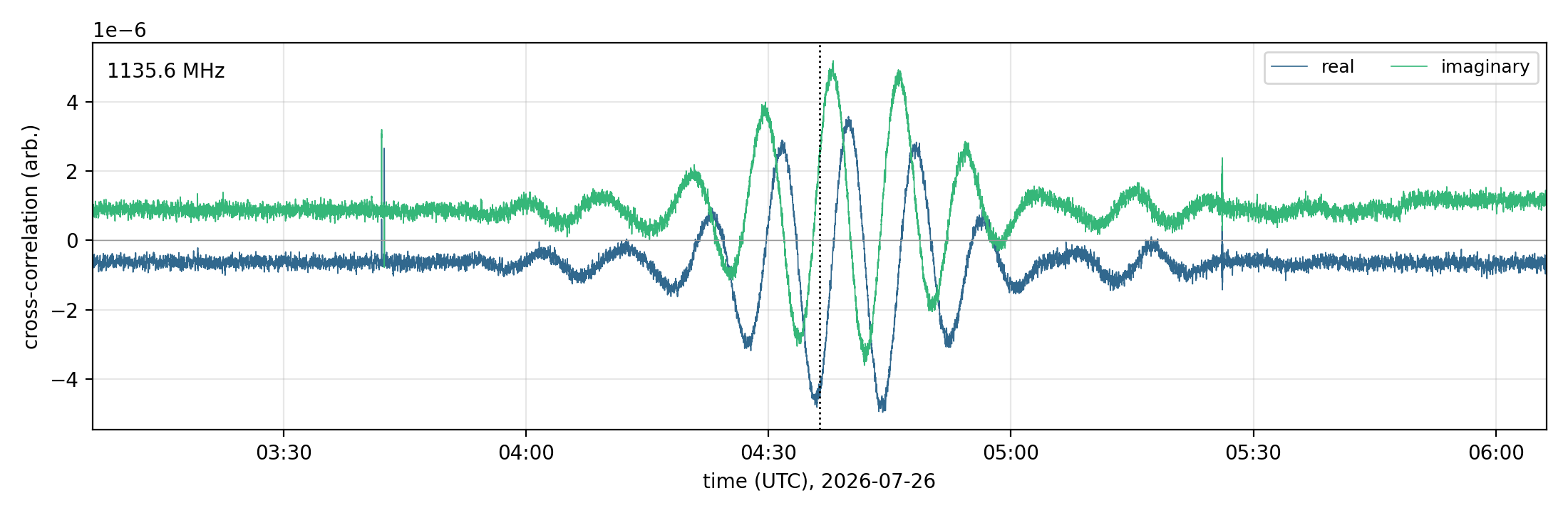}
\caption{Raw (not delay-tracked) cross-correlation at 1135.6\,MHz through the run\#1 \cygA{} transit; the dotted line marks the predicted transit time (04:36:18\,UTC). The real and imaginary parts oscillate in quadrature, with amplitude modulated by the primary-beam crossing. A 2.5\,s boxcar is applied for display; no offset is subtracted.}
\label{fig:cygchan}
\end{figure}

Rather than averaging the cross-correlation across channels (which washes out fringes, since fringe phase depends on frequency for any nonzero interferometer delay), we constructed a time--frequency waterfall of the
raw cross real, imaginary, amplitude, and phase around the run\#1 transit (Figure~\ref{fig:cygfringes}). The result is a clean interferometric delay chirp: the fringe rate across frequency is consistent with passing
through zero at the predicted transit time, within the plot's time resolution, and increases symmetrically away from it in both directions -- the signature of a compact source's geometric delay changing through zero at meridian passage, and independent confirmation that both the correlator and the capture chain preserve coherent cross-correlation phase. The fringe itself is visible without any delay tracking in a single channel (Figure~\ref{fig:cygchan}): at 1135.6\,MHz, the real and imaginary parts of the raw cross-correlation oscillate in quadrature under the transit envelope, peaking at the predicted transit time. The same signature is recovered in the run\#2 window at its own predicted transit time, two days later and at one-fifth of the accumulation length. No quantitative transit-time fit is attempted here; as stated in Section~\ref{sect:intro}, the sky signals are used as validation signals rather than as measurements, and the time axis is in any case
the capture host's packet-arrival timestamp rather than an FPGA-side hardware timestamp.

\subsection{Galactic \texorpdfstring{\HI{}}{HI} plane crossing}
\label{sec:hi}

As a zenith-pointing, non-tracking array, BMX crosses the Galactic plane twice per sidereal day at fixed declination, once toward $l\approx80^\circ$ (through the Cygnus-X complex) and once toward $l\approx166^\circ$ (the outer Galaxy, away from \cygA{} and Cygnus-X). We use the second crossing, at 2026-07-25 13:46\,UTC ($b=-0.001^\circ$), as an independent check that goes beyond the two point-source signals above: rather than a compact source sweeping through the beam, it is an extended, diffuse emission-line signal that must be recovered in autocorrelation power rather than cross-correlation phase or fringing, exercising a different part of the signal chain from Sections~\ref{sec:gnss}--\ref{sec:cygnusa}. Diffuse emission filling the beam is largely resolved out in cross-correlation: at $8.8$\,m the fringe spacing is $1.37^\circ$, so the $4^\circ$ beam contains $\sim$2.9 fringes and smooth structure cancels between them, leaving the autocorrelations as the only products sensitive to it.

Topocentric velocity is computed in the radio convention, $v=c(\nu_0-\nu)/\nu_0$ about the \HI{} rest frequency
($\nu_0=1420.4058$\,MHz), and converted to the kinematic local standard of rest (LSR) frame by adding a single scalar correction (barycentric radial-velocity correction toward zenith, plus the standard 20\,\kms{} solar motion toward the classical apex at RA$=18^{\rm h}$, Dec$=+30^\circ$) evaluated once at the window's midpoint time, not per spectrum. Over the full 5-hour window the barycentric term of this correction drifts from $+25.8$ to $+8.3$\,\kms{}, about one third of the 54.1\,\kms{} channel width; over the $\sim$66-minute on-plane subset that dominates the centroid it drifts by only 4.5\,\kms{}, small compared with the channel width.

\begin{figure}[htbp]
\centering
\includegraphics[width=\linewidth]{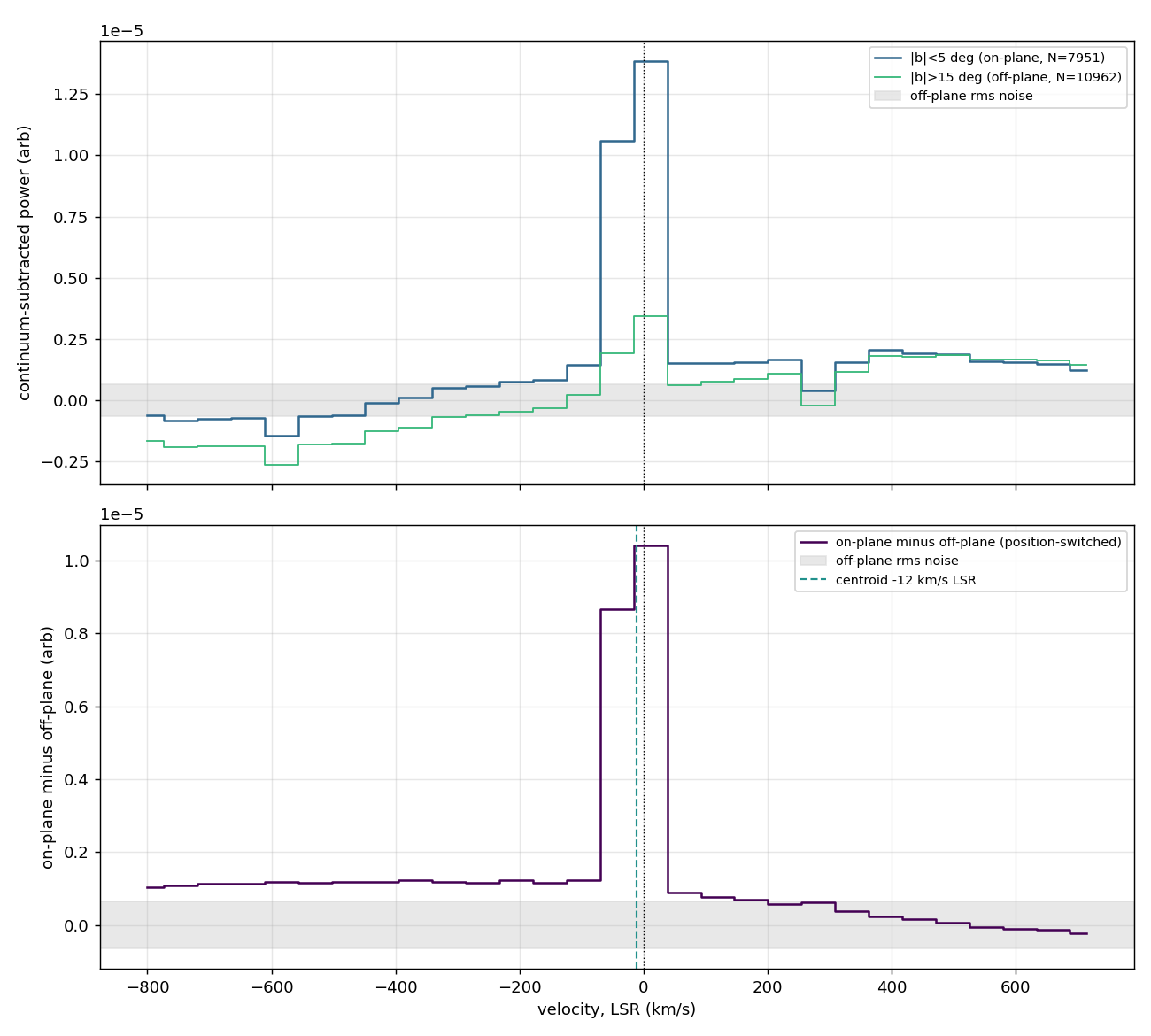}
\caption{Galactic \HI{} 21\,cm plane crossing, run\#1, toward
  $l=166^\circ$. Top: median-stacked, continuum-subtracted
  autocorrelation power vs.\ LSR velocity for the on-plane
  ($|b|<5^\circ$) and off-plane ($|b|>15^\circ$) spectra. Bottom: the
  on-minus-off differential, showing a $16.1\sigma$ peak at the
  intensity-weighted centroid of $-12.0\,\kms$ LSR.}
\label{fig:hi}
\end{figure}

Spectra are split by the zenith beam's instantaneous Galactic latitude into an on-plane set ($|b|<5^\circ$, 7951 spectra) and an off-plane reference set ($|b|>15^\circ$, 10962 spectra); at the $\sim$9$^\circ$\,hr$^{-1}$ rate at which the beam's latitude changes through this crossing, these correspond to $\sim$66\,min on-plane and $\sim$91\,min off-plane. For each spectrum, a second-order polynomial continuum is fit to two flanking baseline windows (1409--1417 and 1426--1436\,MHz) and subtracted before the line window (1417.0--1424.5\,MHz) is extracted. Every on-plane spectrum contains at least one $>10\sigma$ outlier sample within the line search window, consistent with pervasive narrowband RFI in this band; we therefore stack channel-by-channel medians, which are robust to this contamination, and difference the on-plane and off-plane stacks (Figure~\ref{fig:hi}).

\begin{figure}[htbp]
\centering
\includegraphics[width=0.8\linewidth]{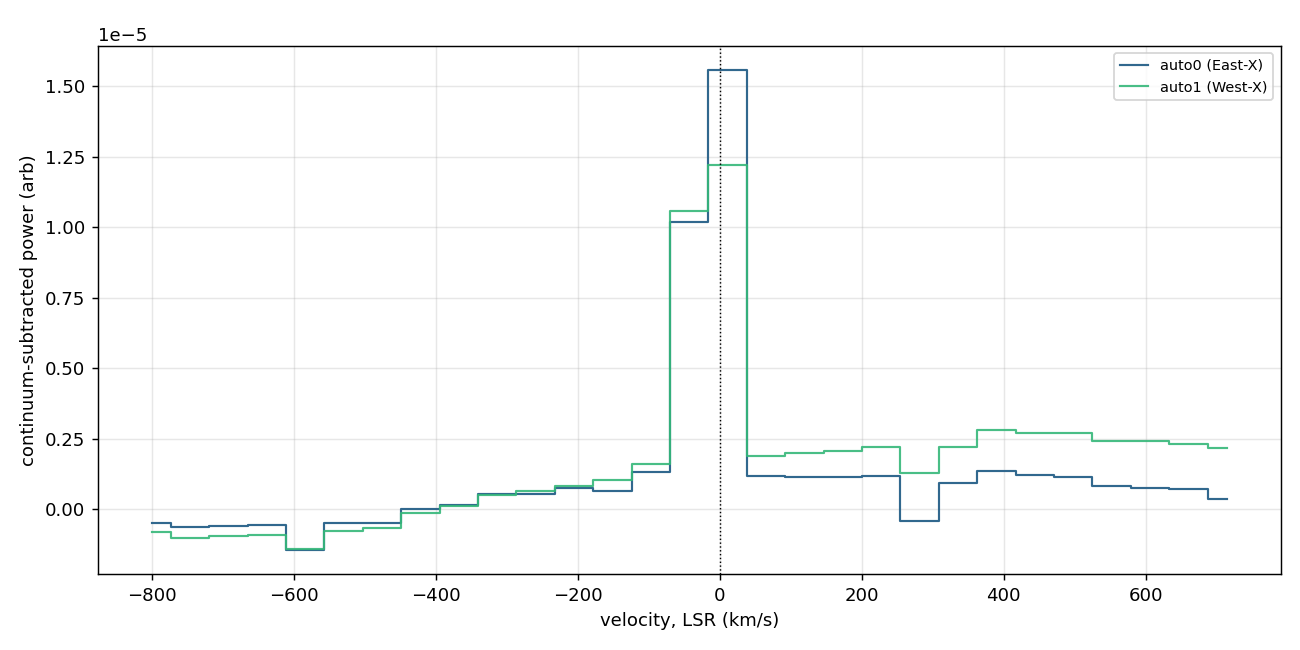}
\caption{The on-plane median-stacked, continuum-subtracted profile of
  Figure~\ref{fig:hi} shown separately for the East (\texttt{auto0}) and
  West (\texttt{auto1}) autocorrelations, supporting the
  independent-dish correlation check in the text.}
\label{fig:hidishes}
\end{figure}

The result is a differential excess coincident with the expected \HI{} velocity, peaking at $16.1\sigma$ above the off-plane noise floor (the per-channel standard deviation across the off-plane spectra, averaged over the line-window channels), with an intensity-weighted centroid of $-12.0\,\kms$ LSR -- physically reasonable for this sightline, close to the Galactic anticenter where the Galactic rotation curve contributes little radial velocity. This significance is a single-bin peak-to-noise ratio computed as if line-window channels were independent; the measured adjacent-channel noise correlation (Section~\ref{sec:radiometer}) is on the order of a few percent, with substantial file-to-file scatter up to $\rho_1\sim0.15$, which would modestly reduce an effective-sample-size-corrected significance below the $16.1\sigma$ headline figure. We have not propagated this into a corrected value here, given the measured correlation's own run-to-run variability, nor have we propagated a formal uncertainty on the centroid via bootstrap or jackknife resampling; the significance should accordingly be read as indicative of a clear, high-confidence detection rather than as a precisely quantified one. The same on/off differencing and median stacking, applied independently to each dish's autocorrelation (Figure~\ref{fig:hidishes}), gives a positive correlation between line-integral power and $-|b|$ of $+0.406$ (East) and $+0.419$ (West) across the full 5-hour window; these are reported as descriptive statistics rather than with formal significance. Two independent receiver chains agreeing on both the sign and approximate magnitude of the effect is evidence against a single-feed instrumental origin, though not against a common-mode signal entering both receivers identically. A broad, low-level residual offset persists between the on- and off-plane stacks across the full search window, plausibly an uncorrected gain or $T_{\rm sys}$ drift over the 5-hour span rather than \HI{}; we therefore restrict the quantitative excess and centroid above
to the narrow feature coincident with the expected line velocity. The profile's second-moment width (143\,\kms, a few channels wide) is reported here only as an upper bound on the true line width given the 54\,\kms{} channel resolution, not as a resolved measurement. As throughout this paper, no absolute flux or brightness-temperature scale is applied; the \HI{} detection is used only as a third, independent confirmation that the backend recovers a real, extended, physically sited sky signal, complementing the compact-source checks above.

\section{Discussion}
\label{sec:discussion}

\subsection{Scaling to the full array}
\label{sec:scaling}

The number of independent correlation products grows quadratically with the number of correlated signal inputs, $P(N)=N^2$ (Eq.~\ref{eq:pn}), so the two-input deployment reported here is the smallest, not the largest, footprint of this design. Throughout this section $N$ counts correlated inputs rather than dishes. The target is $N=8$: both linear polarizations of all four BMX dishes, which uses every RF-ADC on the XCZU48DR. We assume that all products are formed -- eight autocorrelations and all $\binom{8}{2}=28$ cross-correlations, of which 12 are between like polarizations (the set formed by the previous BMX correlator
\cite{oconnor2020}) and 16 between orthogonal polarizations -- giving $P(8)=64$ real quantities per channel. Two parts of the chain scale differently.

\subsubsection{Correlator arithmetic}

The correlator arithmetic is straightforwardly parameterizable: six additional PFB/FFT channelizer chains, an auto-correlator module widened from 2 to 8 power-detector/accumulator pairs, and a cross-correlator module widened from $\binom{2}{2}=1$ to $\binom{8}{2}=28$ complex-multiplier/accumulator pairs, each instantiated identically on the even and odd halves of the spectrum as at present. The RF front end is already provisioned for this: all four ADC tiles are enabled and driven from the common 250\,MHz reference (Section~\ref{sec:rfdc}), and only two of the eight converter channels are presently routed into the DSP chain. Eight inputs span all four tiles, so multi-tile synchronization (MTS), which is not engaged in the present build, must be enabled: without it the relative phase between inputs on different tiles is not deterministic and cross-tile visibilities are not meaningful. The CLK104 already distributes the 5\,MHz SYSREF that MTS requires, but enabling it is a firmware and start-up-sequence change carrying its own bench verification.

The logic footprint at $N=8$ can be bracketed from the measured two-input utilization (Table~\ref{tab:util}). The F-engine (one channelizer per input) scales as $N$, while the X-engine (complex multipliers and vector accumulators) scales as the $P(N)=N^2$ correlation products; bracketing each resource between purely linear ($N/2=4\times$) and purely quadratic ($(N/2)^2=16\times$) growth gives an estimated 28--110\% of CLB LUTs and
14--54\% of DSP slices. The upper ends of these brackets exceed or approach the device, so unlike the buffering below, logic utilization and timing closure at 262.5\,MHz cannot be bounded from the two-input build
and must be established by an implementation run. The true figures lie between the bounds, since only the correlator grows quadratically while the channelizers grow linearly. BRAM is treated separately below.

\subsubsection{Streaming buffer: a calibrated occupancy model}

The visibility packer, FIFO, and serializer are a separate design question. $P(8)=64$ real quantities per channel is sixteen times $P(2)=4$: each channel index now carries 128 accumulated 64-bit fields (64 products on
each spectrum half), or 8192 bits. At the existing 512-bit/262.5\,MHz datapath this requires either an 8192-bit accumulator-to-FIFO interface (with a width converter to re-serialize onto the existing 512-bit packet
datapath) or sixteen sequential 512-bit writes per channel index, dumping $W = N_{\rm chan}/2\times N^2/4 = 16{,}384$ words into the FIFO over 16,384 clocks, against 1024 words over 1024 clocks at $N=2$.

The resulting peak occupancy follows from a simple accounting. Writes arrive at one word per clock for $W$ clocks; the packetizer removes 128 payload words per frame, and a frame admitted at the start of a pacing slot of length $P$ completes 129 clocks later. Writing $n_{\rm c}$ for the number of frames that complete within the write window and $n_{\rm p}$ for words removed by a frame still in progress when writing ends, the peak is
\begin{equation}
  \max(\texttt{dcount}) = W - 128\,n_{\rm c} - n_{\rm p}.
  \label{eq:occupancy}
\end{equation}

We validated Eq.~(\ref{eq:occupancy}) in simulation of the deployed signal chain rather than against an external model. To do so we replaced the fixed 9-bit pacing counter of Section~\ref{sec:pacing} with a 10-bit counter cleared by its own terminal-count comparison against a software register (\texttt{pace\_period}), making the pacing period settable at run time from the control plane without a rebuild. Simulating the unmodified signal chain with this pacer at two settings gives measured peak occupancies of 896 entries at $P=512$ and 490 entries at $P=200$, against predictions of 896 and 488 from Eq.~(\ref{eq:occupancy}); the FIFO returns to exactly zero at the end of every integration in both cases, satisfying the global invariant of
Section~\ref{sec:verification}. The second of these was a genuine out-of-sample prediction. Applied at $N=8$, the same expression gives 12,416 entries at the deployed $P=512$, falling only to 4592 at the 175-clock line-rate floor (Table~\ref{tab:occupancy}): no pacing period brings the $N=8$ dump inside the present 2048-entry FIFO.

\begin{table}[htbp]
\centering
\caption{Peak FIFO occupancy from Eq.~(\ref{eq:occupancy}). Entries marked
$\checkmark$ are values measured in simulation of the deployed firmware
with the run-time pacer (896 and 490, against predictions of 896 and 488);
all other entries are predictions from the same expression. $P$ is the
pacing period in clocks and the inter-frame gap is $P-129$; the line-rate
floor is 175 clocks (Eq.~\ref{eq:tdrain}). The deployed FIFO holds 2048
entries.}
\label{tab:occupancy}
\begin{tabular}{@{}lrrr@{}}
\toprule
$P$ (clk) & Inter-frame gap & $N=2$ ($W{=}1024$) & $N=8$ ($W{=}16384$) \\
\midrule
512 & 383 & 896 $\checkmark$ & 12416 \\
256 & 127 & 640              & 8320  \\
224 &  95 & 512              & 7136  \\
200 &  71 & 490 $\checkmark$ & 6016  \\
176 &  47 & 384              & 4592  \\
\bottomrule
\end{tabular}
\end{table}

\subsubsection{Buffering and BRAM at \texorpdfstring{$N=8$}{N=8}}

The $N=8$ buffer must therefore be enlarged, and what decides feasibility is the total on-chip memory budget, in which the buffer competes with two other structures that grow with $N$: the channelizers (linearly) and the
vector accumulators (quadratically). The per-product readback memories of Section~\ref{sec:correlator} also grow quadratically, but they serve only as a control-plane diagnostic path and are removed in the $N=8$ design,
releasing their BRAM.

Two placements of the enlarged FIFO are available. The XCZU48DR provides, besides its 1080 BRAM tiles, 80 UltraRAM (URAM) blocks (22.5\,Mb), each $4096\times72$\,bit. A FIFO of 16,384 entries -- the full $N=8$ dump, sufficient at any pacing period -- at 512\,bit width occupies $4\times8=32$ URAM blocks (40\%) and no BRAM, leaving the deployed 512-clock pacing period unchanged. URAM is supported as the memory type of the Xilinx parameterized-macro (XPM) synchronous FIFO; if the CASPER FIFO block does not expose that option it can be instantiated as a black box. Alternatively the FIFO can remain in BRAM, at 200 tiles for the 12,416-entry peak at $P=512$, or 72 tiles at $P=176$ if the transmit core is shown to tolerate the shorter inter-frame gap.

Extrapolating the measured two-input utilization (Table~\ref{tab:util}) -- channelizer memory growing linearly in $N$ and vector accumulators quadratically -- gives the budget in Table~\ref{tab:n8bram}. With the readback memories removed, every configuration fits the device; with the FIFO in URAM, BRAM use is 60\% and independent of the pacing period.

\begin{table}[htbp]
\centering
\caption{Estimated on-chip memory at $N=8$ with the per-product readback
memories removed, extrapolated from the measured two-input utilization of
Table~\ref{tab:util} by scaling the channelizer term linearly in $N$ and the
accumulator term as $N^2$. The channelizer term is obtained by difference
and is the dominant and least certain entry; these figures are a bracket,
not a substitute for an implementation run.}
\label{tab:n8bram}
\begin{tabular}{@{}lrrrrrr@{}}
\toprule
FIFO placement & Chan. & Vacc & FIFO & BRAM total & of 1080 & URAM \\
\midrule
URAM, any $P$       & 392 & 256 & --  & 648 & 60\% & 32 of 80 \\
BRAM, $P=512$      & 392 & 256 & 200  & 848 & 79\% & -- \\
BRAM, $P=176$      & 392 & 256 &  72  & 720 & 67\% & -- \\
\bottomrule
\end{tabular}
\end{table}

\subsubsection{Packetization and data volume at \texorpdfstring{$N=8$}{N=8}}

Each spectrum becomes $2N^2=128$ frames. The header's 8-bit total-packet field holds this without modification. The packet-index field is allocated 8 bits in the header but only 3 are driven in the present design (Section~\ref{sec:packetization}), sized for eight packets per spectrum; the driven width must be extended to 7 bits at $N=8$, since left unchanged the index wraps silently and defeats reassembly. Spectrum readout at the deployed 512-clock period becomes $128\times512=65{,}536$ clocks, or 250\,$\mu$s, still negligible against any accumulation period used in this paper.

The output volume grows by the same factor of sixteen, to 128 frames of 8322\,B, or 1.07\,MB, per spectrum. The average link load remains light (85\,Mb\,s$^{-1}$ at $T_{\rm acc}=100$\,ms), but the archived volume becomes $\sim$920\,GB\,day$^{-1}$ at 100\,ms, $\sim$184\,GB\,day$^{-1}$ at 500\,ms, and $\sim$92\,GB\,day$^{-1}$ at 1\,s, so capture-host storage, rather than the link, sets the practical lower limit on accumulation time
for continuous eight-input operation.

\subsubsection{Summary}

We do not claim $N=8$ as demonstrated. The correlator arithmetic and the frame-level packetizer are parameterizable to eight inputs; with the readback memories removed and the visibility FIFO placed in URAM, the
buffering fits the device at the deployed pacing period, with BRAM at an estimated 60\%. Three things remain to be established, and none can be bounded from the two-input build: multi-tile synchronization across all
four ADC tiles, logic utilization, and timing closure at 262.5\,MHz. The most informative next step is an implementation run at $N=8$ -- eight channelizers, the widened correlator, the readback memories removed, and a
URAM FIFO -- which would replace the inferred channelizer term above with a measured one and expose actual logic growth and timing slack. MTS can be commissioned first on two tiles, with four inputs, before it is
relied upon across all four. This two-input deployment is accordingly a commissioned foundation for that extension rather than a demonstration of it.

\subsection{Planned further characterization}
\label{sec:further}

Four items remain to be addressed, and we plan to carry out each of them. The first two cannot be resolved from the archived sky data; the third is newly accessible with the run-time-settable pacer described in Section~\ref{sec:scaling}, and the fourth is a prerequisite for the eight-input design:

\begin{itemize}
\item An injected known-delay test of the cross-correlation sign convention (Table~\ref{tab:word}). The Cauchy--Schwarz check of Section~\ref{sec:cauchyschwarz} validates the visibility word map and transport formats but depends only on cross-correlation \emph{magnitude}; it cannot detect a sign error, such as a swapped East/West label, which an arm-swap step on an injected-delay bench test would.
\item ADC input-level histograms and clipping margin, which cannot be obtained from sky data.
\item A bench sweep of the pacing period over the untested interval between the 175-clock line-rate floor and the deployed 512-clock period (Sections~\ref{sec:pacing} and \ref{sec:scaling}). This would localize the onset of the wire-format corruption, distinguishing a frame-boundary handshake race from sustained oversubscription, and establish the transmit core's minimum inter-frame gap, which sets the required FIFO depth at $N=8$ if the FIFO is kept in BRAM rather than URAM. With the run-time-settable pacer it is a single-bitstream experiment, using the existing sticky FIFO-full flag, transmit-overflow counter and \texttt{acc\_count} continuity as the failure indicators.
\item Commissioning of multi-tile synchronization across all four RF-ADC tiles (Section~\ref{sec:scaling}), including a check that the inter-tile phase relationship is reproducible across power cycles, before cross-tile visibilities are relied upon.
\end{itemize}

\section{Conclusion}
\label{sec:conclusion}

We have implemented a 100\,GbE, ZCU208 RFSoC-based spectrometer/correlator for BMX, realizing the next-generation backend anticipated by O'Connor et al.\cite{oconnor2020}, and documented its digital design end to end: direct RF sampling and channelization, a four-subsystem correlator with BRAM-based vector accumulation, nested timing/synchronization domains, a bit-exact visibility packetization scheme, and a paced packet transmitter, whose necessity we described in detail along with the wire-format corruption it resolves, together with the cycle-accurate testbench used to verify it. Deployed on two independent
multi-hour observing campaigns at two different accumulation times, the system delivers a complete, gap-free visibility stream -- 958,918 consecutive accumulations and 7.67 million packets with zero lost, duplicated, or malformed packets -- independently confirmed end-to-end at the transport-format level by a Cauchy--Schwarz check that is sensitive to, but finds no evidence of, mis-assigned fields in the visibility word map. The channelizer's measured noise performance is consistent with radiometer expectations at $\sim$90--92\% of the
naive prediction, as expected for a polyphase filter bank. The system produces physically sensible sky data: compact-source transits detected and fringe-confirmed across six GNSS bands, the \cygA{} point source detected as a coherent interferometric delay chirp passing through zero at the predicted transit time in both independent campaigns, and a Galactic \HI{} plane crossing detected in autocorrelation and independently reproduced on both dishes. The correlator's arithmetic is algorithmically parameterizable to BMX's full dual-polarization array of eight inputs, and the RF front end is already provisioned for it. At eight inputs the streaming buffer must grow sixteen-fold; we show, against an occupancy model validated in simulation of the deployed firmware at two pacing periods, that with the control-plane readback memories removed and the
visibility FIFO placed in URAM the buffering fits the device at the deployed pacing period, with BRAM at an estimated 60\%. Multi-tile synchronization must additionally be engaged across all four ADC tiles, and logic utilization and timing closure are not bounded by anything measured here. This two-input deployment is accordingly a commissioned foundation for that extension rather than a demonstration of it, and an implementation run at $N=8$ would convert the resource extrapolation above into measurement. With the further characterization outlined in Section~\ref{sec:further}, BMX will be well positioned to use this backend for extended science observing.

\section*{Disclosures}
The authors declare that there are no financial interests, commercial affiliations, or other potential conflicts of interest that could have influenced the objectivity of this research or the writing of this paper.

\section*{Acknowledgments}
BMX was built and optimized with support from the Laboratory Directed Research and Development (LDRD) program at Brookhaven National Laboratory, under projects LDRD~22-070 and LDRD~25-053. The authors thank Tom Throwe and Sam Woronick for IT support. L.P.\ thanks Mitch Burnett, Mayukh Bagchi, Bishnu Sharma, and Rodrigo Rodr\'{i}guez of the CASPER collaboration for useful discussions, and Daniel Keffel, a Summer 2026 Community College Internships (CCI) participant at BNL, for his help with field testing.


\bibliography{report}   
\bibliographystyle{spiejour}   


\vspace{2ex}\noindent\textbf{Liju Philip} is a staff scientist at Brookhaven National Laboratory in Upton, NY, USA. He received his PhD in physics from the University of KwaZulu-Natal, South Africa, in 2019. His research interests include radio astronomy instrumentation, RFSoC-based digital signal processing, and cosmology, as well as hardware development and system integration for underwater neutrino experiments.

\listoffigures
\listoftables

\end{spacing}
\end{document}